\documentclass[reprint,aps,prd,superscriptaddress]{revtex4-2}
\usepackage{amsmath}
\allowdisplaybreaks[4]
\usepackage{multirow}
\usepackage{amsfonts}
\usepackage{xcolor}
\usepackage{comment}
\usepackage{graphicx}
\usepackage{siunitx}
\usepackage{ulem}
\usepackage{booktabs}
\usepackage[
    pdfauthor={Teng Zhang},
    pdftitle={ET-room temperature},
    colorlinks=true,
    linkcolor=black,
    urlcolor=blue,
    citecolor=blue
]{hyperref}

\DeclareSIUnit{\sqrthz}{\ensuremath{\sqrt{\text{\hertz}}}}

\begin{document}
\renewcommand\texteuro{FIXME}
\allowdisplaybreaks[4]


\title{A Broadband Room-Temperature Design for the Einstein Telescope}

\author{Teng Zhang}
\affiliation{School of Physics and Astronomy, and Institute for Gravitational Wave Astronomy, University of Birmingham, Edgbaston, Birmingham B15\,2TT, United Kingdom}

\author{Amit Singh Ubhi}
\affiliation{School of Physics and Astronomy, and Institute for Gravitational Wave Astronomy, University of Birmingham, Edgbaston, Birmingham B15\,2TT, United Kingdom}

\author{Huan Yang}
\affiliation{Department of Astronomy, Tsinghua University, Beijing 100084, China}

\author{Alan V. Cumming}
\affiliation{Institute for Gravitational Wave Research, School of Physics and Astronomy, University of Glasgow, Glasgow G12 8QQ, United Kingdom}

\author{Mark Barton}
\affiliation{Institute for Gravitational Wave Research, School of Physics and Astronomy, University of Glasgow, Glasgow G12 8QQ, United Kingdom}

\author{Angus Bell}
\affiliation{Institute for Gravitational Wave Research, School of Physics and Astronomy, University of Glasgow, Glasgow G12 8QQ, United Kingdom}

\author{Fabio Bergamin}
\affiliation{Gravity Exploration Institute, Cardiff University, Cardiff CF24 3AA, United Kingdom}

\author{Giacomo Bruno}
\affiliation{Centre for Cosmology, Particle Physics and Phenomenology (CP3), Institute of Research in Mathematics and Physics (IRMP), Université catholique de Louvain (UCLouvain), Louvain-la-Neuve, Belgium}

\author{Katherine L. Dooley}
\affiliation{Gravity Exploration Institute, Cardiff University, Cardiff CF24 3AA, United Kingdom}

\author{Johannes Eichholz}
\affiliation{School of Physics and Astronomy, and Institute for Gravitational Wave Astronomy, University of Birmingham, Edgbaston, Birmingham B15\,2TT, United Kingdom}

\author{Mariana Fazio}
\affiliation{SUPA, University of Strathclyde, Glasgow G1 1XQ, UK}

\author{Aaron Goodwin-Jones}
\affiliation{Centre for Cosmology, Particle Physics and Phenomenology (CP3), Université catholique de Louvain, Chemin du Cyclotron 2, 1348 Louvain-la-Neuve, Belgium}

\author{James Hough}
\affiliation{Institute for Gravitational Wave Research, School of Physics and Astronomy, University of Glasgow, Glasgow G12 8QQ, United Kingdom}

\author{Keiko Kokeyama}
\affiliation{Gravity Exploration Institute, Cardiff University, Cardiff CF24 3AA, United Kingdom}

\author{Haixing Miao}
\affiliation{Frontier Science Center for Quantum Information, Department of Physics, Tsinghua University, Beijing, 100084, China}

\author{Xingrui Peng}
\affiliation{School of Physics and Astronomy, and Institute for Gravitational Wave Astronomy, University of Birmingham, Edgbaston, Birmingham B15\,2TT, United Kingdom}

\author{Geraint Pratten}
\affiliation{School of Physics and Astronomy, and Institute for Gravitational Wave Astronomy, University of Birmingham, Edgbaston, Birmingham B15\,2TT, United Kingdom}

\author{Stuart Reid}
\affiliation{SUPA, University of Strathclyde, Glasgow G1 1XQ, UK}

\author{Sheila Rowan}
\affiliation{Institute for Gravitational Wave Research, School of Physics and Astronomy, University of Glasgow, Glasgow G12 8QQ, United Kingdom}

\author{Patricia Schmidt}
\affiliation{School of Physics and Astronomy, and Institute for Gravitational Wave Astronomy, University of Birmingham, Edgbaston, Birmingham B15\,2TT, United Kingdom}

\author{Alberto Vecchio}
\email{a.vecchio@bham.ac.uk}
\affiliation{School of Physics and Astronomy, and Institute for Gravitational Wave Astronomy, University of Birmingham, Edgbaston, Birmingham B15\,2TT, United Kingdom}

\author{Giles Hammond}
\email{Giles.Hammond@glasgow.ac.uk}
\affiliation{Institute for Gravitational Wave Research, School of Physics and Astronomy, University of Glasgow, Glasgow G12 8QQ, United Kingdom}

\author{Hartmut Grote}
\email{hartmut.grote@astro.cf.ac.uk}
\affiliation{Gravity Exploration Institute, Cardiff University, Cardiff CF24 3AA, United Kingdom}

\author{Denis Martynov}
\email{d.martynov@bham.ac.uk}
\affiliation{School of Physics and Astronomy, and Institute for Gravitational Wave Astronomy, University of Birmingham, Edgbaston, Birmingham B15\,2TT, United Kingdom}

\begin{abstract}

The unique underground location of the Einstein Telescope (ET) is expected to substantially suppress seismic and gravity-gradient noise, providing a favorable environment for gravitational-wave detection at low frequencies compared with surface-based observatories. The baseline ET design aims to achieve broadband sensitivity through a xylophone configuration, in which separate cryogenic and room-temperature interferometers are optimised for low- and high-frequency gravitational-wave detection. However, achieving the target low-frequency sensitivity introduces substantial technical complexity caused by vertical suspension thermal noise, vibrations of the cryocooler, and detuned optical configuration. In this paper, we demonstrate an alternative, entirely room-temperature, fully resonant, single-interferometer architecture that addresses these challenges and achieves broadband sensitivity superior to that of the conventional ET xylophone configuration. Using mature fused-silica suspension technology, recent advances in optical coatings, and novel inertial isolation, we show that 4-m-long fused-silica fibres can meet the required suspension thermal-noise performance. We further demonstrate that vibrations, optical rigidity, and quantum back-action noise can be controlled through appropriate design of the interferometer, test mass suspensions, optical seismometers, and quantum states.
\end{abstract}

\maketitle

\section{Introduction}

The Einstein Telescope (ET) is a next-generation gravitational-wave observatory designed to achieve an order-of-magnitude improvement in broadband strain sensitivity over current detectors\,\cite{Maggiore2020ScienceCase,Hild_2011, ETDesignReport2020, Branchesi2023ETDesigns, Punturo2011ThirdGeneration}. In particular, extending sensitivity to lower frequencies is crucial for detecting more massive black-hole binaries at high redshifts\,\cite{gair2011exploring} and for observing low-mass compact binaries earlier in their inspiral. Access to the early inspiral would enable earlier warnings of imminent mergers\,\cite{PhysRevD.97.123014} and provide longer signal baselines for measuring subtle binary dynamics, including spin precession and orbital eccentricity\,\cite{PhysRevResearch.2.043096,PhysRevD.105.023003}. Improved low-frequency sensitivity is therefore one of the central scientific drivers of the Einstein Telescope.

The baseline ET xylophone architecture employs separate high-power, room-temperature and low-power, cryogenic interferometers optimized for high- and low-frequency detection, respectively~\cite{Hild_2010,Hild_2011}. This approach comes at the cost of substantial technical complexity associated with cryogenic operation, including heat extraction~\cite{Koroveshi2023CryogenicPayloads}, molecular adsorption and icing of cold optics~\cite{Hasegawa2019MolecularAdsorbedLayer}, and additional noise mechanisms associated with silicon test masses~\cite{Siegel2023TCCR,Klochkov2022ElectricFieldSilicon}. Moreover, the ET-LF interferometer employs a detuned signal-recycling configuration~\cite{PhysRevD.67.062002}, introducing additional optomechanical couplings and sensing and control challenges~\cite{Ward2010}. Experience with LIGO and Virgo~\cite{Capote2025AdvancedLIGO, Virgo2026Advanced} shows that the measured detector noise can exceed the modelled fundamental-noise budget because of technical noise sources that are difficult to fully characterise. This gap between modelled and realised sensitivity may pose an even greater challenge for ET-LF, where the extreme low-frequency sensitivity requirements demand simultaneous control of numerous mechanical, optical, thermal, seismic, and control-noise mechanisms.

\begin{figure}[t]
    \centering
    \includegraphics[width=\columnwidth]{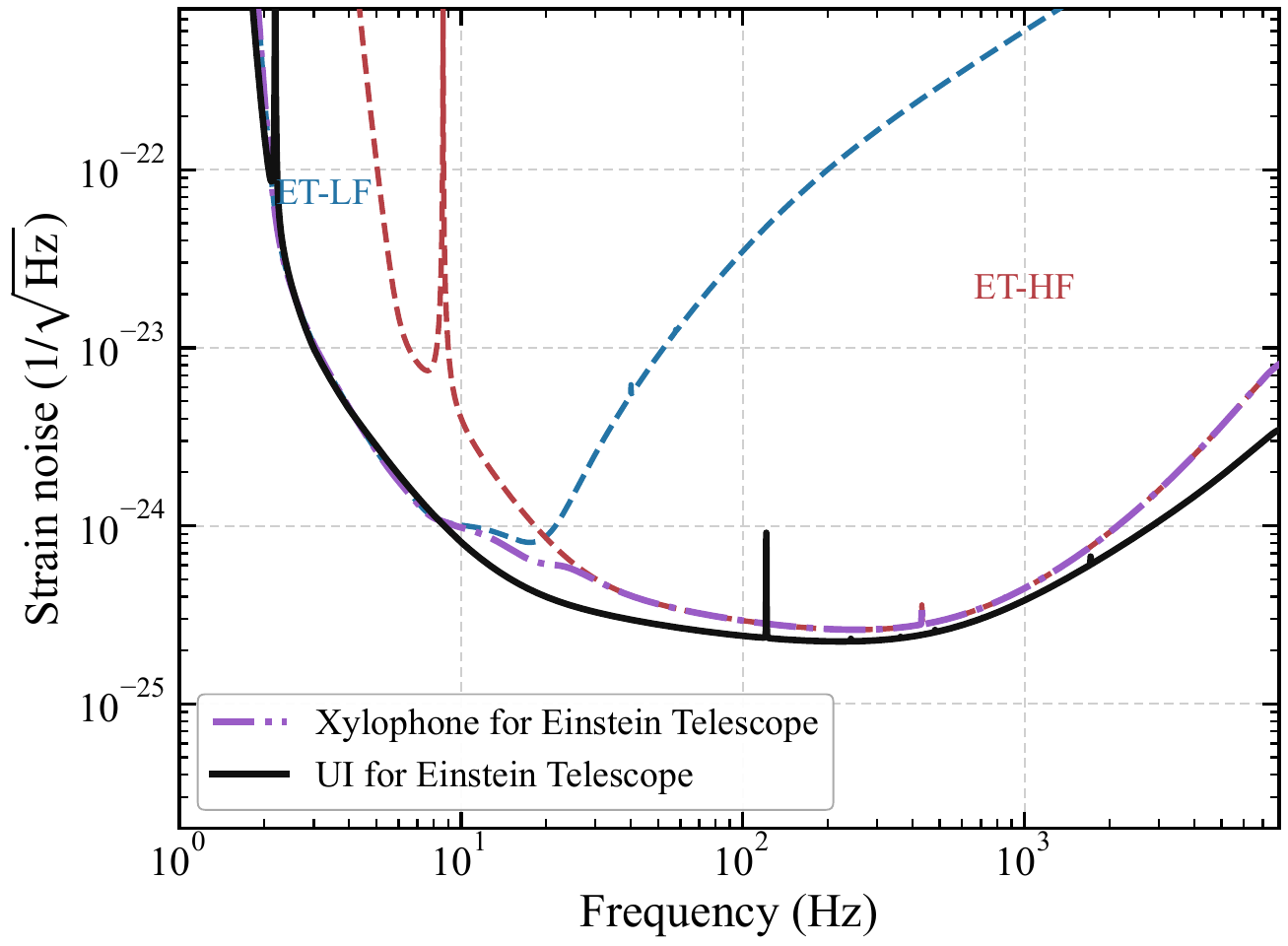}
    \caption{Sensitivities of the proposed UI design for ET, a
\SI{15}{\kilo\meter} broadband room-temperature detector and the
\SI{15}{\kilo\meter}  baseline
xylophone configuration consisting of ET-LF and ET-HF interferometers. The HF and LF detector sensitivities of the Xylophone configuration are illustrated by red and blue dashed lines. The sensitivities correspond to antenna response  for a source at a zenith angle of
$15^\circ$.
}
    \label{fig:total-budget}
\end{figure}

\begin{figure}[t]
    \centering
\includegraphics[width=\columnwidth]{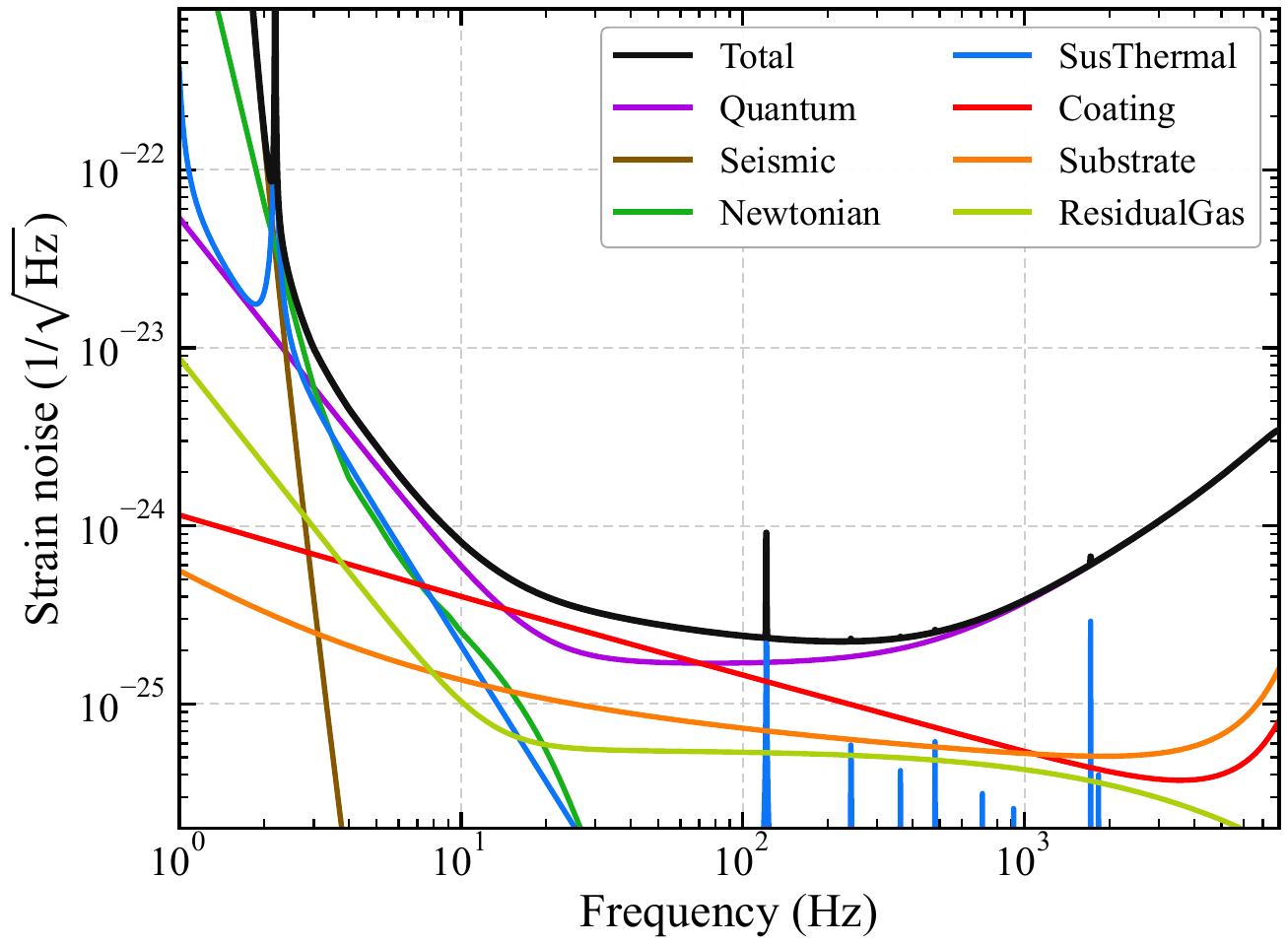}
    \caption{Noise budget of the UI design for ET, a
\SI{15}{\kilo\meter} broadband room-temperature detector. The thermal noise of the proposed fused silica suspension matches the gravity-gradient noise, which is estimated at the quiet Sos Enattos site as a representative example.}
    \label{fig:rt-budget}
\end{figure}

These considerations motivate the exploration of a detector architecture with fewer coupled subsystems and less stringent technical requirements, such as a unified, room-temperature interferometer that can provide broadband sensitivity while avoiding both cryogenic operation and signal-recycling detuning. We refer to this concept as the unified interferometer (UI) for ET. Figures~\ref{fig:total-budget} and~\ref{fig:rt-budget} show the resulting broadband sensitivity and its noise budget, respectively. The parameter set used in the calculation is summarized in Table~\ref{tab:design-summary}. The low-frequency performance is enabled by \SI{440}{\kilogram} test masses, which reduce the displacement response to radiation pressure, and $\SI{4}{\meter}$-long, high-stress final-stage fused-silica fibres, which suppress suspension thermal noise through enhanced dissipation dilution. A \SI{5}{\kilo\meter} filter cavity enables low-loss frequency-dependent squeezing to be maintained over the detection band.

In the following sections, we first discuss the design of the fused-silica suspension and the technological solutions adopted to mitigate vertical suspension thermal noise~\cite{PenaArellano2025VerticalThermalNoise}. We then examine the control challenges of the UI and demonstrate how 440\,kg test masses mitigate high-power effects such as optical rigidity. Next, we present the broadband and kHz-optimized sensitivity of the detector and investigate its scaling with arm length. Finally, we compare the resulting performance with that of Cosmic Explorer.

\begin{table*}[t]
\caption{Principal parameters of the proposed \SI{15}{\kilo\meter} room-temperature detector for ET.}
\label{tab:design-summary}
\begin{ruledtabular}
\begin{tabular}{ll}
Parameter & Proposed value \\
\hline
\multicolumn{2}{l}{\textit{Interferometer configuration}} \\
\hline
Arm length & \SI{15}{\kilo\meter} \\
Laser wavelength & \SI{1064}{\nano\meter} \\
Input laser power & \SI{500}{\watt} \\
Arm circulating power & \SI{3}{\mega\watt} \\
ITM transmissivity & \SI{1.4}{\percent} \\
ETM transmissivity & $5\,\mathrm{ppm}$ \\
ITM/ETM radius of curvature & \SI{7651}{\meter} \\
Signal-extraction mirror (SEM) transmissivity, broadband/high frequency & \SI{8}{\percent} / \SI{1}{\percent} \\
Signal-extraction-cavity (SEC) length & \SI{100}{\meter} \\
\hline
\multicolumn{2}{l}{\textit{Squeezing and optical losses}} \\
\hline
Injected squeezing & $18\,\mathrm{dB}$ \\
Observed squeezing & $\sim 10\,\mathrm{dB}$\\
Injected squeezing angle & \SI{0}{\radian} \\
Filter-cavity length & \SI{5}{\kilo\meter} \\
Arm loss per mirror & $37.5\,\mathrm{ppm}$  \\
Signal-extraction-cavity loss & $500\,\mathrm{ppm}$ \\
OPA-cavity / injection Faraday-isolator loss & \SI{1}{\percent} / \SI{2}{\percent} \\
Photodetector / Faraday-isolator / output-mode-cleaner loss & \SI{1}{\percent} / \SI{1}{\percent} / \SI{1}{\percent} \\
Filter-cavity end-mirror transmissivity / round-trip loss & $5\,\mathrm{ppm}$ / $60\,\mathrm{ppm}$ \\
Squeezer phase noise & \SI{10}{\milli\radian}  \\
Filter-cavity / signal-extraction-cavity length noise (rms) & \SI{1}{\pico\meter} / \SI{1}{\pico\meter} \\
\hline
\multicolumn{2}{l}{\textit{Test masses, beam, and suspension}} \\
\hline
Test-mass mass & \SI{440}{\kilogram} \\
Test-mass temperature & \SI{290}{\kelvin} \\
Test-mass radius / thickness & \SI{37}{\centi\meter} / \SI{46.5}{\centi\meter} \\
Test-mass substrate & Fused silica \\
Beam radius on the test masses & \SI{13.5}{\centi\meter} \\
Coating loss angle & $5\times10^{-5}$/$1.3\times10^{-5}$ \\
Final suspension-fibre material & Fused silica \\
Number of final-stage fibres & 4 \\
Final suspension-fibre length & \SI{4}{\meter} \\
Final suspension-fibre neck length & \SI{0.03}{\meter}\\
Final suspension-fibre stress & \SI{2}{\giga\pascal} \\
Fibre surface loss & $2.5\times 10^{-12}$m\\
\end{tabular}
\end{ruledtabular}
\end{table*}

\section{Test mass suspension}
\label{sec:sus}

A key component of the UI for ET is its room-temperature test-mass suspension. Our aim is to achieve the suspension thermal noise~\cite{Gabriela2000} at or below the level of the ET facility limit given by the gravity-gradient noises at low frequencies. For the seismic input, we use the underground seismic spectrum measured at Sos Enattos as representative of a seismically quiet underground environment~\cite{DiGiovanni2020SosEnattos}.
The gravity-gradient noise arises from the direct gravitational coupling of environmental density fluctuations to the test masses.
In this study, we retain the ET model for the Rayleigh-wave, body-wave, atmospheric, and cavern-acoustic contributions\,\cite{harms2015terrestrial,amann2020site,harms2022lower, Janssens2022CorrelatedET}. We assume that local seismic and acoustic sensor arrays reduce the amplitude spectral densities of the corresponding Newtonian-noise contributions by a factor of three.

The noise level of the gravity gradients, shown in Fig.~\ref{fig:rt-budget}, is at the level of $\approx 2.5 \times 10^{-25}$\,${\rm strain}/{\sqrt{\rm Hz}}$ and also sets the required limit on the suspension thermal noise of $3.75\times 10^{-21}$\,${\rm m}/\sqrt{\rm Hz}$. In this section, we propose a solution to the ET suspension thermal noise problem, covering both the bottom suspension stage and a well-known problem of the vertical suspension thermal noise coming from the metal blades of the final filter stage~\cite{Baratti2025ETLFVerticalThermalNoise}. In this section, we show that \SI{4}{\meter}-long fused-silica fibres can achieve the required suspension thermal-noise performance.

\subsection{Bottom stage suspension thermal noise}

Gains in suspension thermal-noise performance from the final suspension stage arise directly from the deployment of three basic concepts, as previously outlined in~\cite{Cumming_22}. First, a larger test mass reduces the thermal noise, with the relevant scaling inversely proportional to the square root of the test-mass mass, according to the fluctuation-dissipation theorem~\cite{Saulson_90, PhysRev.83.34_FDT, PPP2009, Gabriela2000}. Second, longer suspension fibres reduce the resonant frequency of the fundamental pendulum mode, thereby reducing the off-resonance thermal noise in the detection band of the instrument. For the \SI{4}{\meter}-long fibres proposed here, the final-stage pendulum frequency is reduced to \SI{0.246}{\hertz}, compared with \SI{0.642}{\hertz} for the \SI{0.6}{\meter}-long fibres used in the LIGO suspensions. Longer suspensions also provide additional dissipation-dilution benefits by reducing the ratio of elastic bending energy to gravitational energy.

The scaling of the horizontal suspension thermal noise with fibre length, \(L_{\rm fibre}\), and test-mass mass, \(M_{\rm tm}\), relative to the LIGO-estimated value~\cite{Cumming_2020}, is given by
\begin{equation}
\sqrt{S_{xx}(10\,{\rm Hz})} \approx 7 \times 10^{-20} \frac{\rm m}{\sqrt{\rm Hz}} \times \frac{0.6\,{\rm m}}{L_{\rm fibre}} \times \sqrt{\frac{40\,{\rm kg}}{M_{\rm tm}}}.
\end{equation}

Since the LIGO horizontal suspension thermal noise is approximately a factor of 19 above the estimated gravity-gradient noise level at the Sos Enattos facility, we need to achieve a comparable reduction in suspension thermal noise by increasing the fibre length and test-mass mass. Choosing \(L_{\rm fibre} = 4\,\mathrm{m}\) and \(M_{\rm tm} = 440\,\mathrm{kg}\), we obtain a reduction in the thermal noise of approximately a factor of 22 relative to the LIGO horizontal suspension noise. This estimate does not require any reduction in the critical fibre surface loss.


\begin{figure}[t]
    \centering
\includegraphics[width=\columnwidth]{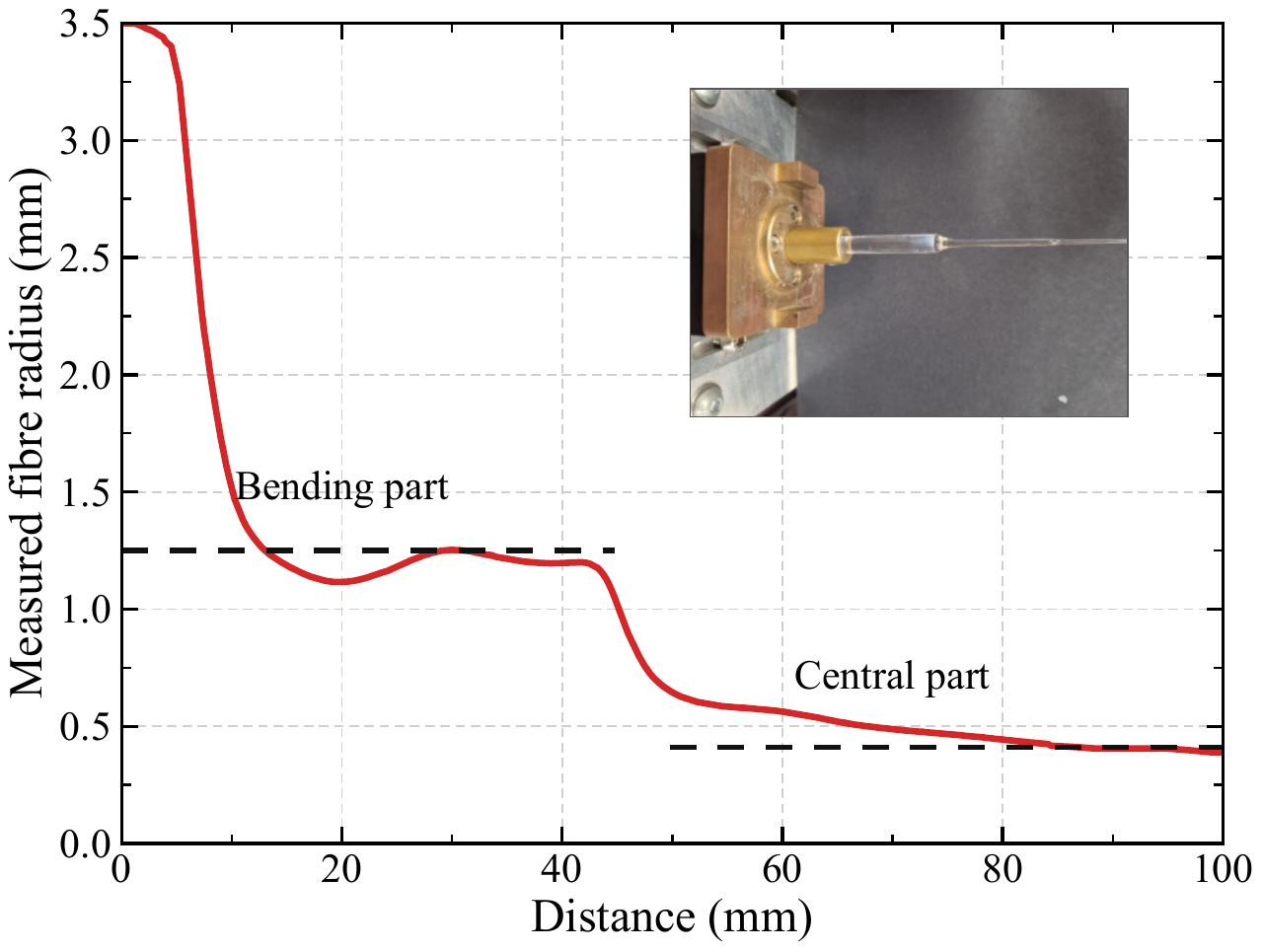}
    \caption{Example of a measured fibre profile from the University of Glasgow's CE fibre pulling tests.}
    \label{fig:fibre-profile}
\end{figure}

The fibre design proposed here follows the standard, well-established approach in which the otherwise dominant thermoelastic noise contribution is minimised by exploiting a particular property of fused silica: its Young's modulus increases with temperature, i.e., the material stiffens as the temperature rises~\cite{Cagnoli_02}. This effect can be balanced against the thermal expansion coefficient, allowing the thermoelastic dissipation to be completely cancelled for a specific tensile stress in the fibre's bending regions, namely \SI{185}{\mega\pascal}. For a \SI{440}{\kilogram} test mass, this corresponds to a fibre diameter of \SI{2728}{\micro\metre} in the bending regions. A length of \SI{30}{\milli\metre} at each end of the fibre is specified at this diameter, which is sufficient to accommodate the bending of the fibre.

In addition to horizontal suspension thermal noise, vertical suspension thermal noise can couple into the GW readout channel because, for a long-baseline detector, the local gravity direction cannot be exactly perpendicular to the arm direction due to the curvature of the Earth. The corresponding minimum geometrical coupling of vertical test-mass motion into the GW channel is
\begin{equation}
\theta = \frac{L}{2R_E} \approx 1.2 \times 10^{-3},
\end{equation}
where \(L\) is the detector arm length and \(R_E\) is the Earth's radius.

The central section of the fibre is operated at a higher stress, $\sigma$, allowing the suspension's vertical bounce and roll frequencies to be reduced and ultimately shifted out of the detection band for $\sigma \gtrsim 1$\,GPa, and also providing passive filtering of the vertical suspension thermal noise from the final filter. The higher stress also increases the frequencies of the individual fibre violin modes. 
The scaling of the vertical suspension thermal noise coupled to the horizontal readout for a 15\,km-baseline detector, relative to the LIGO-estimated value~\cite{Cumming_2020}, is given by
\begin{equation}
\begin{split}
\sqrt{S_{vv}(10\,{\rm Hz})} & \approx 7 \times 10^{-21} \frac{\rm m}{\sqrt{\rm Hz}} 
\times \sqrt{\frac{0.6\,{\rm m}}{L_{\rm fibre}}} \\
&
 \times \left({\frac{40\,{\rm kg}}{M_{\rm tm}}} \right)^\frac{3}{4} \times \left(\frac{0.77\,{\rm GPa}}{\sigma} \right)^\frac{1}{4}
\end{split}
\end{equation}
and is significantly below the required suspension thermal noise level for the fibre stress values of $\sigma = 1-2$\,GPa.

\begin{figure}[t]
    \centering
\includegraphics[width=\columnwidth]{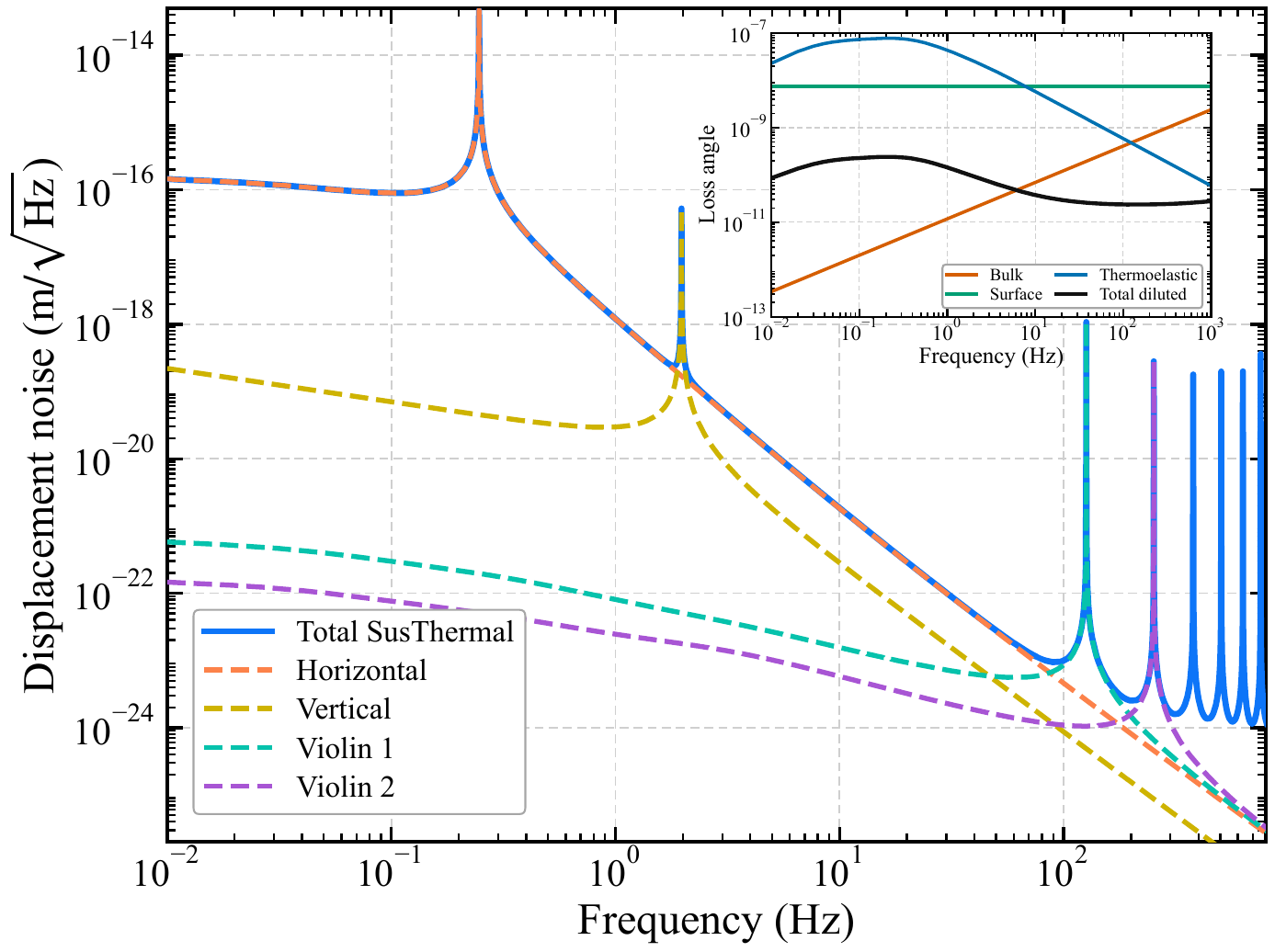}
    \caption{Suspension thermal noise budget of a single test mass from FEA modelling. The vertical-to-longitudinal coupling coefficient is taken as ${L}/{2R_E}$. Inset: Mechanical loss angles associated with the different dissipation mechanisms in the considered fibre. The measured fibre profile places the crossover between thermoelastic and surface losses at \(7.7\,\mathrm{Hz}\).}
    \label{fig:sus-budget}
\end{figure}

Suspensions supporting larger masses at higher stresses have already been demonstrated and provide a viable basis for further development~\cite{Cumming_22, Acernese2023}. Ongoing research into stress corrosion in fused-silica fibres~\cite{lee2019suspension, Toland2020} 
suggests that a practical current limit on the ultimate fibre stress is approximately \SI{2}{\giga\pascal}, corresponding to a strength safety factor of approximately 2.5. Previous studies have also found no clear correlation between fibre length and tensile failure strength, suggesting that the increased fibre length does not in itself represent a significant additional risk. For a \SI{440}{\kilogram} test mass and four suspension fibres, the stress requires a reduction in the fibre cross-sectional area, giving a central fibre diameter of \SI{828}{\micro\metre}.

Manufacturing longer fibres is relatively straightforward, requiring only an extension of the pulling apparatus~\cite{Heptonstall_11}. Suspension fibres with lengths of up to \SI{1.2}{\meter} have already been successfully produced~\cite{Cumming_22}, and no fundamental difficulties are anticipated in extending the process to \SI{4}{\meter}.

Fibres with the required profile have already been successfully demonstrated  
using \SI{7}{\milli\metre}-diameter fused-silica stock as the starting material, as shown in Fig.~\ref{fig:fibre-profile}. This enabled the full fibre geometry to be realised, including accurate modelling of the neck profiles at the transitions between different diameters. The experimentally observed variation in diameter achievable during the fibre-pulling process was also incorporated through the thermoelastic-nulling region, as shown in Fig.~\ref{fig:fibre-profile}. This level of realism is included from the outset in both the design methodology and the thermal-noise modelling.

Suspension thermal noise is modelled using finite element analysis (FEA).
The FEA model accurately models the elastic bending energy in the system, allowing accurate calculation of the contributing noise components, together with the dissipation dilution of the pendulum. This permits an advanced level of fidelity of the noise model to be included in the proposed suspension design. Suspension thermal noise is calculated using the FEA-based methodologies extensively detailed in~\cite{Cumming_2009, Cumming_2012}, with the FEA model calculating the individual loss components surface, bulk, and thermoelastic- that contribute mechanical dissipation to each low-frequency resonant mode of the pendulum system. The loss components are shown in the inset of  Fig.~\ref{fig:sus-budget}.

The projected single test mass thermal noise is shown in Fig.~\ref{fig:sus-budget}, with the projected displacement noise being $1.85 \times 10^{-21}$\,${\rm m}/\sqrt{\rm Hz}$ at 10\,Hz, which translates to a strain sensitivity in a \SI{15}{\kilo\meter} detector of $2.46 \times 10^{-25}$\,${\rm strain}/\sqrt{\rm Hz}$ at 10 Hz. 

\subsection{Vertical suspension thermal noise from the final filter}

\begin{figure}[t]
    \centering
    \includegraphics[width=0.6\columnwidth]{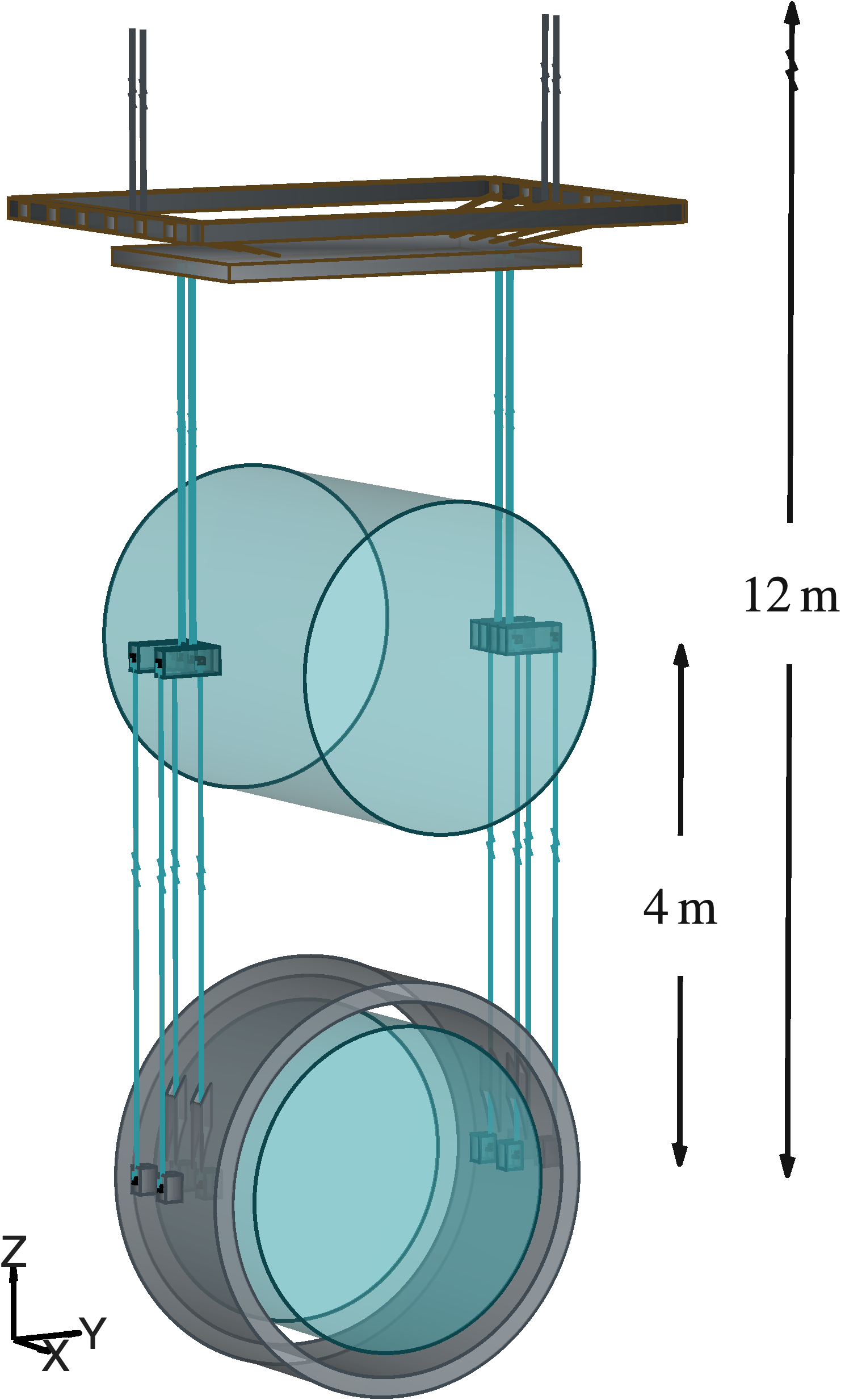}
    \caption{Schematic of the test mass, reaction mass, marionette, and final filter stage considered in this section.}
    \label{fig:TestMass}
\end{figure}


The proposed long fused-silica fibres also provide passive filtering of suspension thermal noise from the upper suspension stages, which have relatively high metal loss angles of \(\sim 10^{-3}\)~\cite{Braccini2000Maraging}. In particular, the vertical springs in the multi-stage suspension attenuate the coupling of ground vibrations to the test mass. However, the final vertical filter is itself a significant source of suspension thermal noise~\cite{Gennai2023ReverseEngineering}. This is a particular limitation for the low-frequency sensitivity of ET-LF: thermal noise generated by the final vertical filter is transmitted to the test mass and can couple into horizontal motion through the curvature of the Earth~\cite{Baratti2025ETLFVerticalThermalNoise}. Silicon flexures at the two bottom stages of ET-LF suspensions can inject negligible vertical thermal noise, but do not provide the required filtering of the blades' thermal noise below 15-25\,Hz.


\begin{figure}[t]
    \centering
\includegraphics[width=\columnwidth]{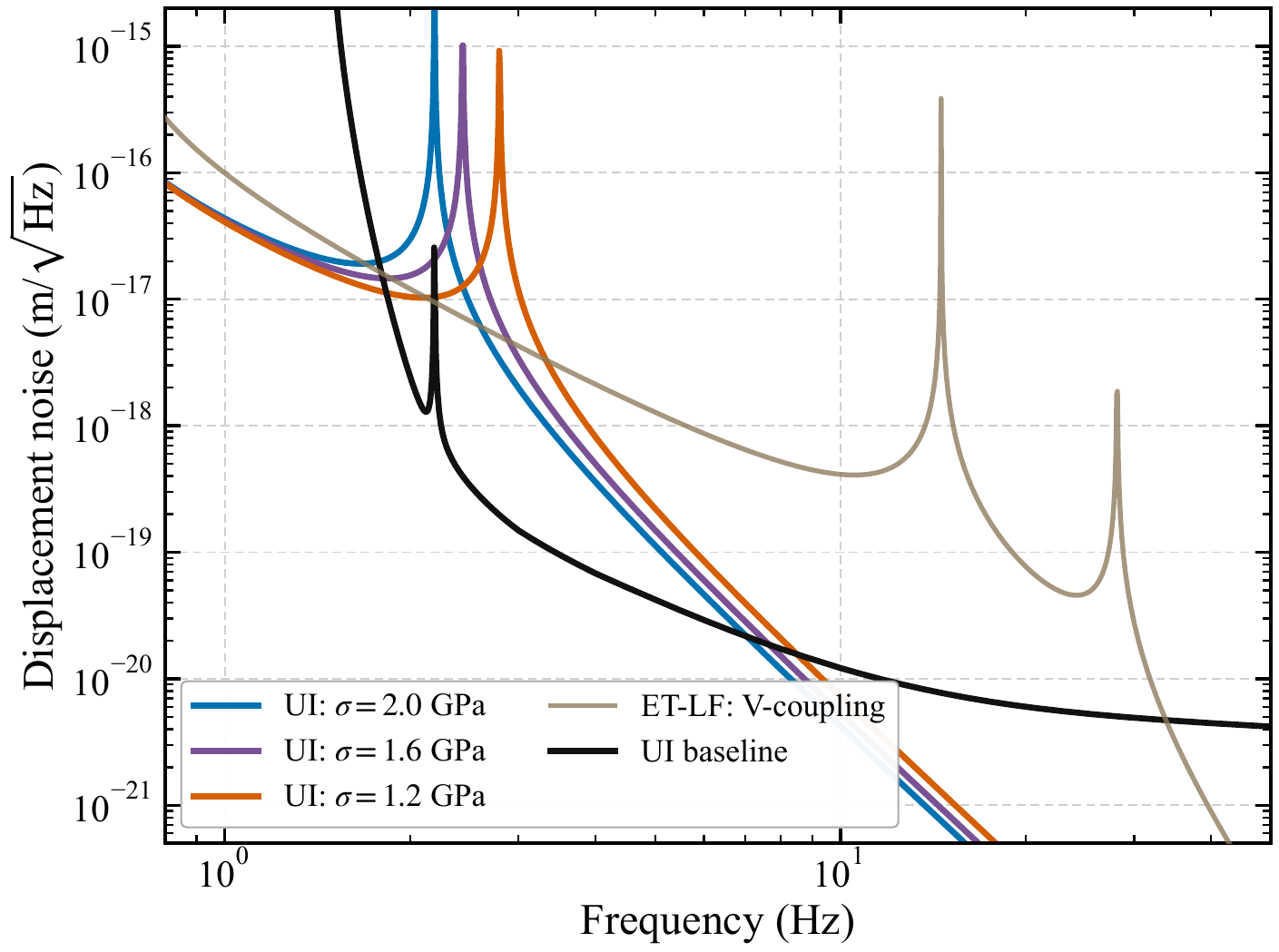}
    \caption{Filtering of the vertical suspension thermal noise by a 4-m-long fused-silica fibre with a stress of $\sigma = 1.2 - 2$\,GPa. The figure also illustrates the benefit of passive filtering compared with the rigid-body modes of the ET-LF vertical suspension~\cite{PenaArellano2025VerticalThermalNoise}. The thermal noise shown is the combined contribution from all four test masses.
    }
    \label{fig:sus_thermal_v}
\end{figure}

Long fused-silica fibres between the test mass and the marionette in the UI concept address this challenge. Fused silica combines very low mechanical loss with a sufficiently low vertical stiffness and supreme mechanical strength to provide a passive filtering stage. In our design, the bounce mode is at $\approx 2$\,Hz, providing a $1/f^2$ attenuation of vertical motion above this frequency, as shown in Fig.~\ref{fig:sus_thermal_v} for the fibre stress of $\sigma = 1.2 - 2$\,GPa. The simulation includes a conventional final filter with maraging-steel blades, a vertical eigenfrequency of 0.5\,Hz, and a Q-factor of $\approx 64$, based on KAGRA data~\cite{Baratti2025ETLFVerticalThermalNoise} and measurements of the mechanical loss of maraging steel~\cite{Braccini2000Maraging}. The resulting vertical thermal noise provides the reference against which the filtering performance of the long fused-silica fibre is evaluated.

The vertical suspension thermal-noise coupling to the GW channel estimated in Fig.~\ref{fig:sus_thermal_v} should be regarded as a conservative estimate, assuming no further advances in the design of the vertical springs or sensing system. Over the ET development timescale, this contribution could potentially be reduced further through optimisation of the vertical suspension. Below, we consider three options of reducing the vertical suspension thermal noise below the baseline UI sensitivity.

\subsubsection{Fused silica vertical blades}

Possible improvements of vertical suspension thermal noise include increasing the length of the maraging-steel blades, thereby reducing their mechanical stiffness, or developing vertical blades from lower-loss materials such as fused silica~\cite{LIGO2020InstrumentScienceWhitePaper}, sapphire~\cite{Kumar2016KAGRA, Hammond2018QuasiMonolithic}, or silicon~\cite{Adhikari2020CryogenicSilicon}.

\begin{figure}[t]
    \centering
\includegraphics[width=\columnwidth]{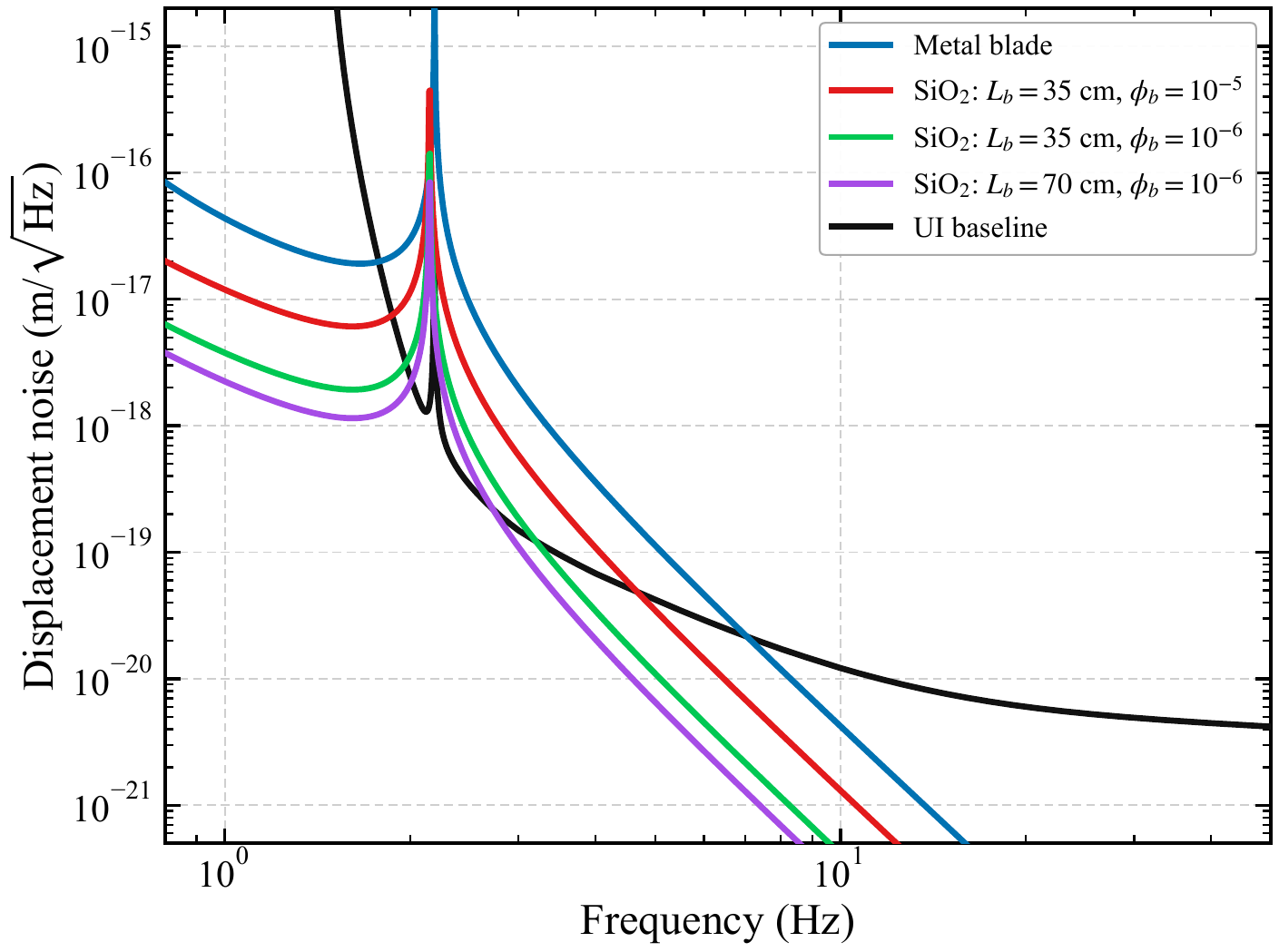}
    \caption{Vertical thermal noise from the final filter stage with fused-silica blades. The blade thickness is tuned to maintain a constant blade stress of \(77\,\mathrm{MPa}\), regardless of the blade length.}
    \label{fig:sus_thermal_v_blades}
\end{figure}

Fused silica is a natural choice for the vertical blades of the room-temperature UI suspension due to its exceptionally low mechanical dissipation. We estimate that the thermoelastic, surface, and bulk losses of a fused-silica blade are each \(<10^{-8}\), such that the mechanical dissipation is expected to be dominated by clamping losses. In this section, we estimate the vertical suspension thermal noise associated with the fused-silica blades required to support the marionette, test-mass, and reaction masses, which have a combined mass of \(M_{\rm tot}\approx1760\,\mathrm{kg}\). Assuming 12 fused-silica blades, the load carried by each blade is $ F_b = M_{\rm tot}g / 12 \approx1.44\,\mathrm{kN}.$

For a LIGO-type triangular blade with length \(L_b\), clamped-end width \(w_b\), and thickness \(t_b\), the vertical resonance frequency can be estimated from the static deflection \(\delta\) under the applied load,
\begin{equation}
\omega_v^2 \approx \frac{g}{\delta} \approx \frac{g Y w_b t_b^3}{6 F_b L_b^3} = \frac{Y}{\sigma_b}\frac{t_b}{L_b}\frac{g}{L_b},
\end{equation}
where $\sigma_b$ is the bending stress and is given by
\begin{equation}
\sigma_b =
\frac{6F_bL_b}{w_bt_b^2}.
\end{equation}
We choose the blade dimensions to keep \(\sigma_b\leq 80\,\mathrm{MPa}\) and to achieve a margin against mechanical failure.

Realistic geometrical parameters for the fused-silica blades result in a relatively stiff vertical mode of the final filter, with
$\omega_v\approx2\pi\times(3.5-6.5)\,\mathrm{Hz}.$
This mode can subsequently be softened to approximately \(0.3\,\mathrm{Hz}\) using magnetic anti-springs, following a concept similar to that employed in the current Superattenuator design. While this softening does not reduce the intrinsic thermal noise of the blades, it can improve the isolation of the test mass from vertical suspension thermal noise generated in the higher suspension stages. Total vertical suspension thermal noise coupling to the beam direction is shown in Fig.~\ref{fig:sus_thermal_v_blades} for $w_b = 10$\,cm.

\subsubsection{Local sensing of the test mass}
Another approach to subtracting vertical suspension thermal noise is the use of local sensing~\cite{PenaArellano2025VerticalThermalNoise}. If asymmetries in the fibre stress prove challenging to control, the long fused-silica fibres in the UI concept provide an additional opportunity to reduce this coupling for symmetric suspensions. Since the bounce mode is at approximately 2\,Hz, the test mass is effectively inertial with respect to the marionette at frequencies above the bounce resonance. This enables a direct measurement of the relative vertical motion between the marionette and the test mass, analogous to the measurement principles employed in inertial sensors and gravimeters. The measured relative motion can therefore be used to estimate the vertical suspension thermal noise and construct a feedforward signal for subtraction from the GW channel, either in real time or during post-processing.

\begin{figure}[t]
    \centering
\includegraphics[width=\columnwidth]{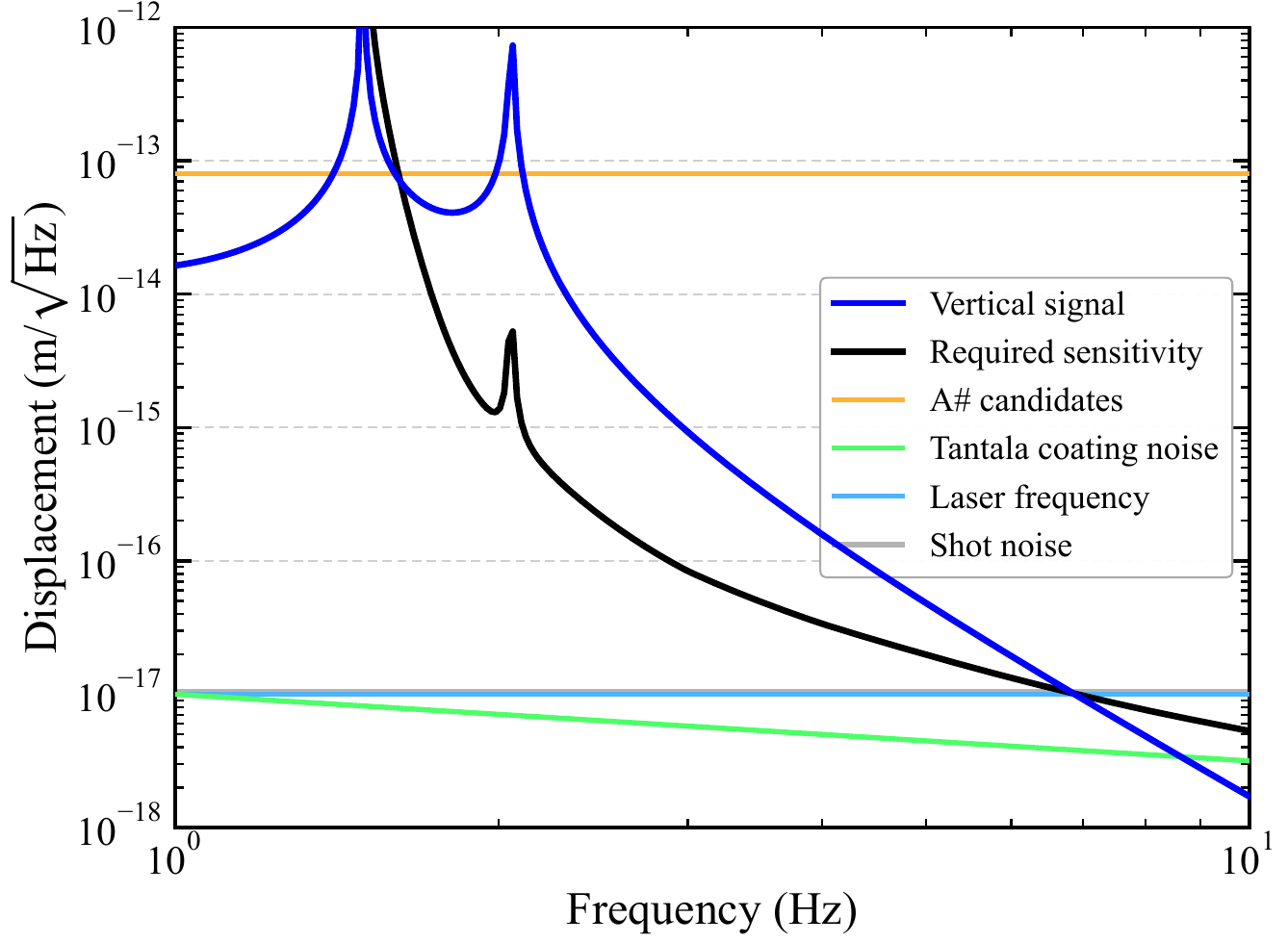}
    \caption{Required sensitivity and analysis of fundamental noise sources of compact interferometers required to measure the relative motion between the marionette and test masses.}
    \label{fig:sus_thermal_v_sens}
\end{figure}

The required sensitivity of the interferometric sensor measuring the relative motion between the marionette and the test mass is shown in Fig.~\ref{fig:sus_thermal_v_sens}. State-of-the-art compact Michelson interferometers currently achieve sensitivities of approximately $0.03$--$0.2$\,pm/$\sqrt{\mathrm{Hz}}$ in the 1--10\,Hz frequency band~\cite{6DPRL, Cooper2018HoQI, Isleif2019CompactMultifringe}. Fabry--Pérot interferometers with stabilised lasers offer the potential to improve the sensitivity further, reaching the required level of approximately $0.01$--$0.1$\,fm/$\sqrt{\mathrm{Hz}}$ over the 2--7\,Hz frequency band.

\subsubsection{Vertical to pitch coupling}
\begin{figure}[t]
    \centering
\includegraphics[width=\columnwidth]{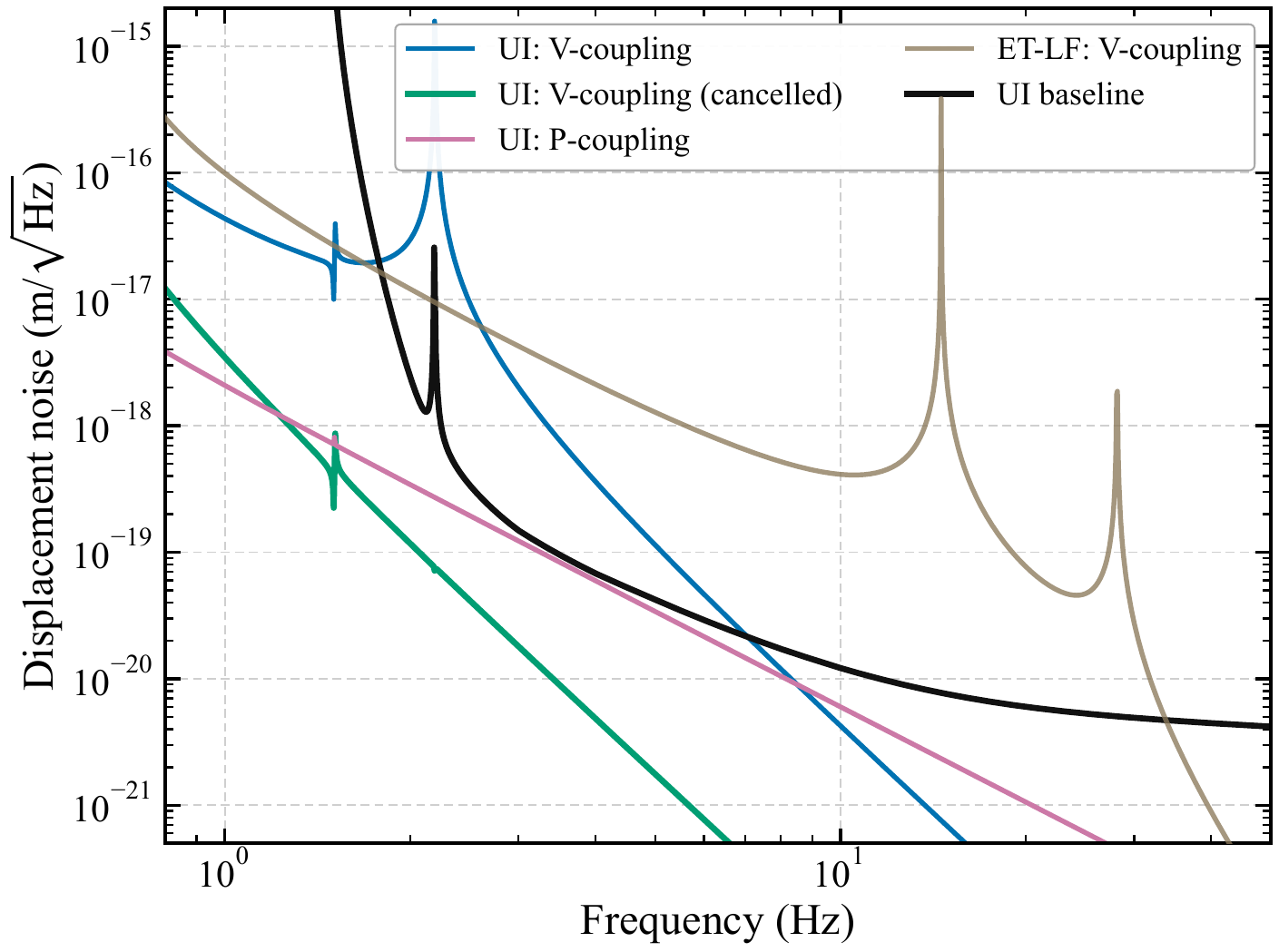}
    \caption{
    The figure shows the suppression of the vertical coupling to horizontal displacement (V-coupling) achieved through an intentional asymmetry in the fibres, together with the corresponding thermal-noise coupling from pitch to horizontal displacement (P-coupling).
    }
    \label{fig:sus_thermal_v_to_p}
\end{figure}

In addition to passive filtering, long fused-silica fibres may further suppress the coupling of vertical suspension thermal noise into the horizontal direction. We propose to achieve this by introducing a 20\% asymmetry in the stress between the front and back test-mass fibres. This asymmetry introduces an additional coupling between the vertical motion of the marionette and the pitch motion of the test mass. By appropriately positioning the arm-cavity beam on the test mass relative to the pitch rotation axis, this coupling can be used to cancel the vertical suspension thermal noise, as shown in Fig.~\ref{fig:sus_thermal_v_to_p}. For our chosen set of parameters, the required beam offset is 7.5\,mm. We note that the cancellation of suspension thermal noise through coupling between orthogonal degrees of freedom has been extensively studied in the literature~\cite{Braginsky1999,Wipf2013}.

The vertical offset of the beam position relative to the pitch axis also introduces coupling of the test-mass pitch thermal noise into the horizontal degree of freedom sensed by the beam. However, this additional contribution, shown as P-coupling in Fig.~\ref{fig:sus_thermal_v_to_p}, remains below the baseline sensitivity because fused silica is a low-loss material. We also note that the fibre asymmetry introduces coupling between the thermal pitch motion of the marionette and the vertical motion of the test mass. We consider this contribution to be negligible, however, because the marionette can be suspended from the final filter stage using a short fused-silica fibre and we leave this contribution for further investigations.

The proposed fibre asymmetry is compatible with current fused-silica fibre manufacturing capabilities~\cite{Heptonstall_11}. The required difference in fibre length is approximately 22\,mm, while the present pulling process is expected to achieve a length accuracy of approximately 0.2\,mm, with substantially higher-resolution length encoders available for further improvement. Similarly, the required difference in the central fibre radius is approximately 40\,$\mu$m. Although current production tolerances are based on a 5\% acceptance criterion, improved control of the fibre-pulling process and stabilisation of the heating region make reproducible production of fibres with the required diameters achievable\,\cite{Lee_2019}. Some batch production and selection would likely still be required, as is already standard practice, but we do not anticipate this to present a significant additional manufacturing challenge.

\subsection{Suspension inside the vacuum tank}

The preliminary ET Superattenuator design~\cite{Bertocco2024NGSA} demonstrates the feasibility of accommodating a multi-stage suspension within an overall length of approximately \(12\,\mathrm{m}\). A comparable suspension height is adopted in the ET baseline detector layout~\cite{ETBaselineDetectorLayout2025}. This height provides a useful geometrical constraint for the UI design considered here: the complete suspension system, including the intermediate attenuation stages and the final suspension of the test mass, must be accommodated within the available vacuum-tank height. In particular, adopting a \(4\,\mathrm{m}\)-long final fibre requires the remaining suspension stages to be arranged within the remaining envelope. 
In this paper, we analyse the seismic noise coupling to the GW channel by considering a vibration isolation system that consists of an active platform~\cite{Matichard2015LIGOISI1, Matichard2015LIGOISI2} and a six-stage suspension.

\begin{figure}[t]
    \centering
    \includegraphics[width=\columnwidth]{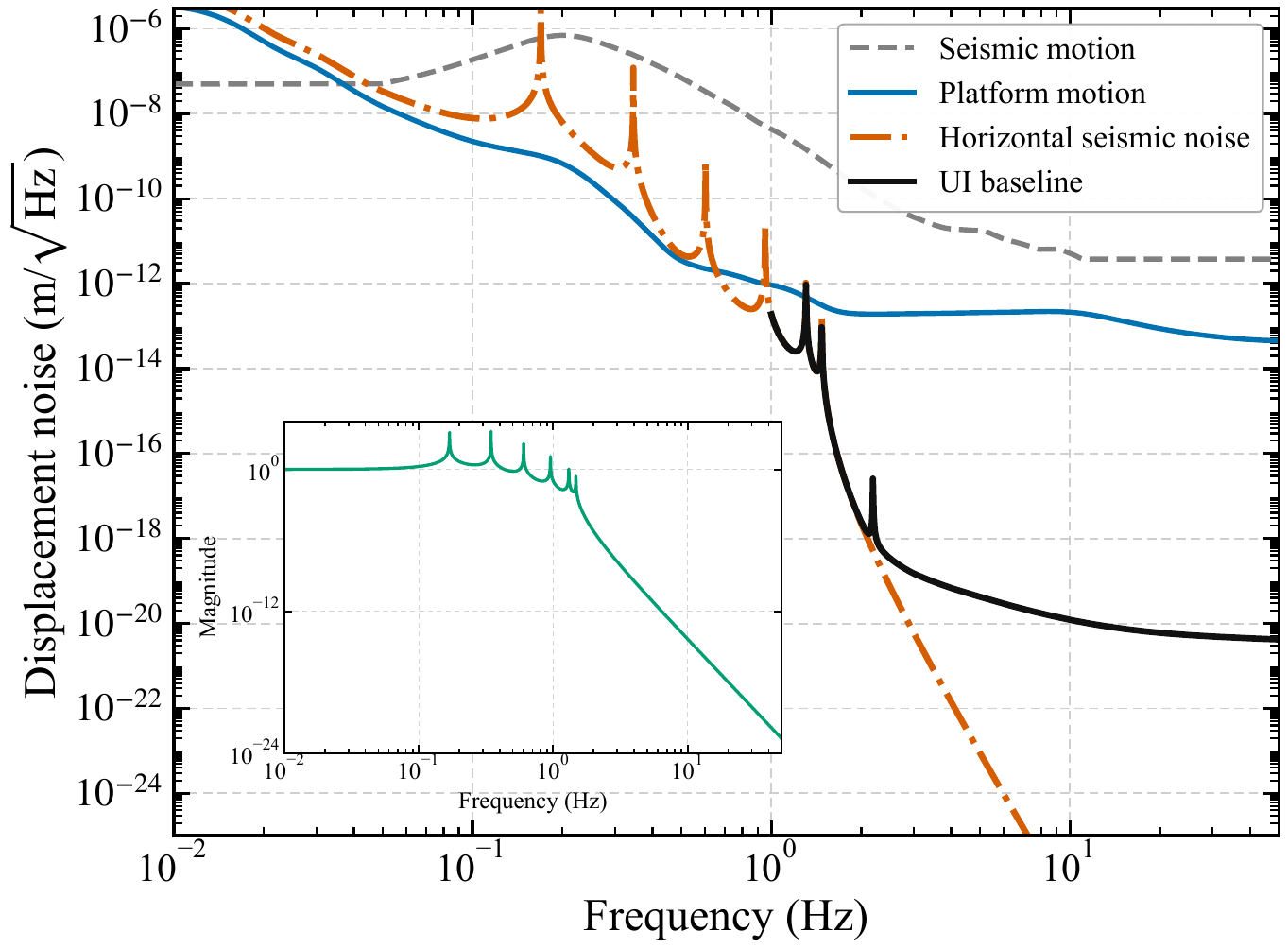}
    \caption{Simulated residual platform motion assuming sensing with a Compact 6D seismometer, compared with the extrapolated seismic motion at Sos Enattos site. The resulting horizontal test-mass noise is shown together with the UI sensitivity. Inset: the magnitude of suspension transfer function from platform motion along the beam axis to the longitudinal test-mass response.}
    \label{fig:ET_UI_seismic}
\end{figure}

The platform performance adopted in the noise budget is shown in Fig.~\ref{fig:ET_UI_seismic}. The platform utilises state-of-the-art optical seismometers and commercial geophones for inertial stabilisation starting from 10\,mHz and provides at least two orders of magnitude of suppression of the support motion across the principal suspension resonances in the \SIrange{0.1}{1}{\hertz} band\,\cite{6Dproposal,6Dproto,6Dsilica,6Dplatform,6DPRL}.


We propose a six-stage pendulum suspension designed to provide sufficient passive seismic isolation to meet the UI seismic-noise requirements above \(3\,\mathrm{Hz}\). The suspension comprises four suspended filter stages followed by the marionette and the test mass, with the upper four pendulum stages and the marionette stage having lengths of \(1.6\,\mathrm{m}\), and a \(4\,\mathrm{m}\)-long final suspension fibre between the marionette and the test mass. The total length of the wires and fibres at all stages is 12\,m. The resulting coupled mechanical system is modelled as a simplified six-degree-of-freedom pendulum. We leave other degrees of freedom for future studies.

Simulations of the resulting seismic transfer function indicate that the proposed configuration can reduce the coupling of ground motion to the test mass by a factor of $\approx 5 \times 10^7$. This approach is therefore particularly attractive for a design constrained by the available \(12\,\mathrm{m}\) vacuum-tank height, while retaining a \(4\,\mathrm{m}\) final suspension fibre to provide enhanced low-frequency isolation. The result of the simulation is shown in Fig.~\ref{fig:ET_UI_seismic}.

\section{Quantum and classical control}

The proposed interferometer operates on resonance and avoids detunings in the signal recycling cavity because they cause auxiliary degrees of freedom and technical noises to couple into the GW readout. The detuning also complicates the interferometer's sensing and control because it modifies the optical response and introduces optical-spring dynamics, resulting in a more strongly coupled length-control system and making lock acquisition and stable operation more challenging ~\cite{Ward2010}. Furthermore, the resulting coupled control system is more sensitive to parameter variations and makes real-time antenna calibration challenging. Finally, detuning rotates the squeezing quadrature in a frequency-dependent manner, complicating the injection of frequency-dependent squeezed vacuum~\cite{Peng2024}.

The choice of an on-resonance broadband interferometer requires $\sim$MW of circulating power in the arms to achieve the baseline sensitivity of UI, both at low and high frequencies. In this section, we discuss the requirements for the quantum states, test-mass geometry, and inertial isolation required to achieve the baseline sensitivity.

\subsection{Squeezed states of light}

\begin{figure}[t]
    \centering
\includegraphics[width=\columnwidth]{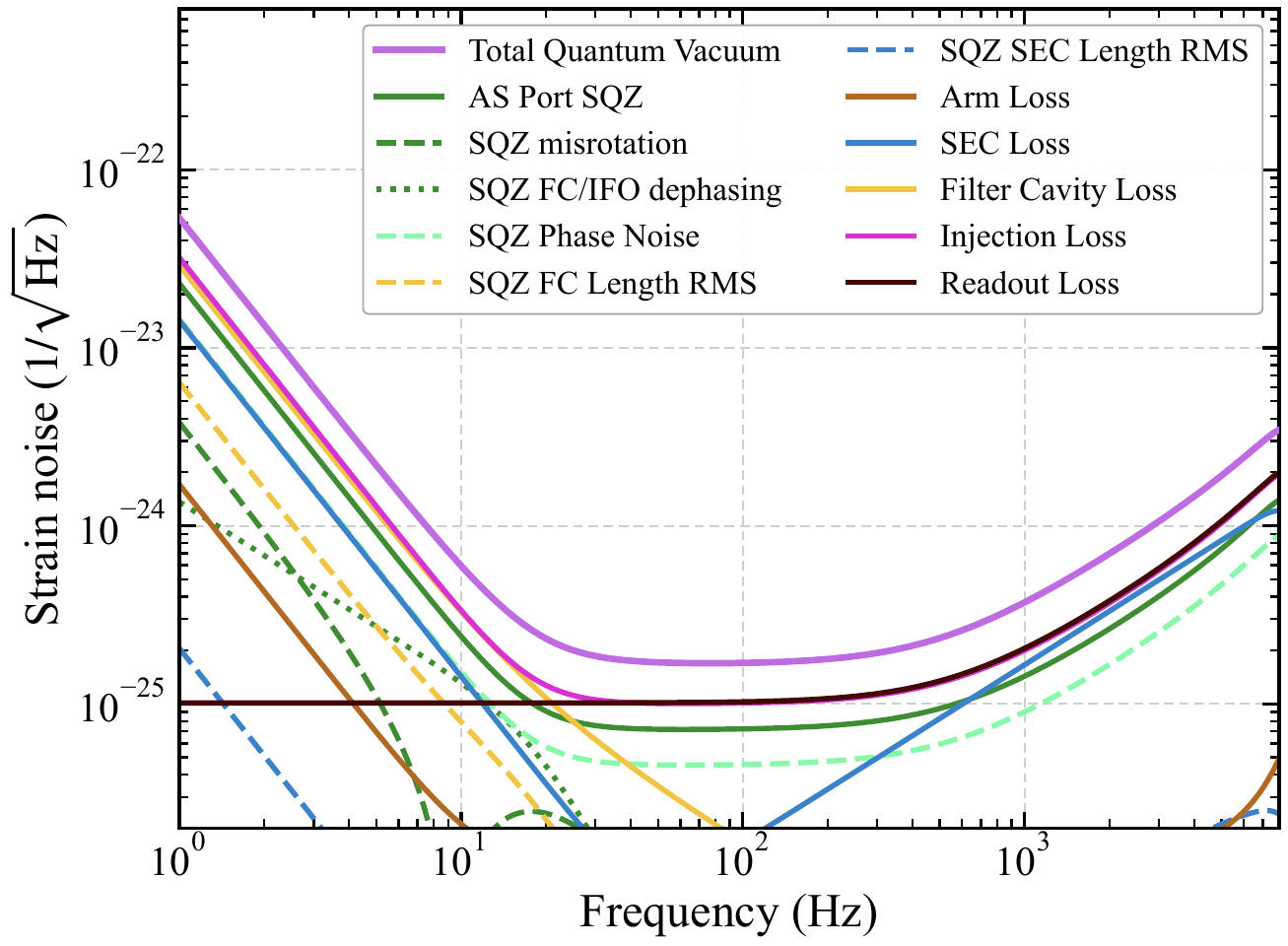}
    \caption{Quantum noise budget of the \SI{15}{\kilo\meter} broadband UI room-temperature detector.}
    \label{fig:QN}
\end{figure}

Quantum noise has two limiting contributions. At high frequencies, photon-counting statistics give rise to shot noise in the phase quadrature, while at low frequencies, amplitude fluctuations of the optical field exert fluctuating radiation-pressure forces on the test masses. Increasing the circulating arm power reduces shot noise but increases radiation-pressure noise. Injecting squeezed states of light with a properly tuned filter cavity suppresses both noise terms~\cite{Aasi_13, Grote_13, Acernese_23, Barsotti_19, Jia_24}.

The proposed design relies on low optical losses that preserve the frequency-dependent squeezed vacuum injected at the detector output.
The optical losses of the filter cavity are particularly important at low frequencies. For a given bandwidth required to achieve the desired quadrature rotation, a longer filter cavity can operate at lower finesse
\,\cite{PhysRevD.88.022002,PhysRevD.98.022010} and reduce the total effective optical loss. We therefore adopt a single \SI{5}{\kilo\meter}-long filter cavity for the proposed UI design. By comparison, the baseline ET-LF model considered here employs two detuned \SI{1}{\kilo\meter}-long filter cavities. Although the UI filter cavity is longer, the use of a single cavity avoids the need to accommodate two parallel optical paths and their associated vacuum and control systems. This reduces the required tunnel width and simplifies both the underground infrastructure and the optical configuration. We show the detailed quantum noise budget in Fig.~\ref{fig:QN}, with relevant parameters listed in Table~\ref{tab:design-summary}.

\subsection{Optical rigidity}\label{sec:OpticalRigidity}

\begin{figure}[t]
    \centering
    \includegraphics[width=1\columnwidth]{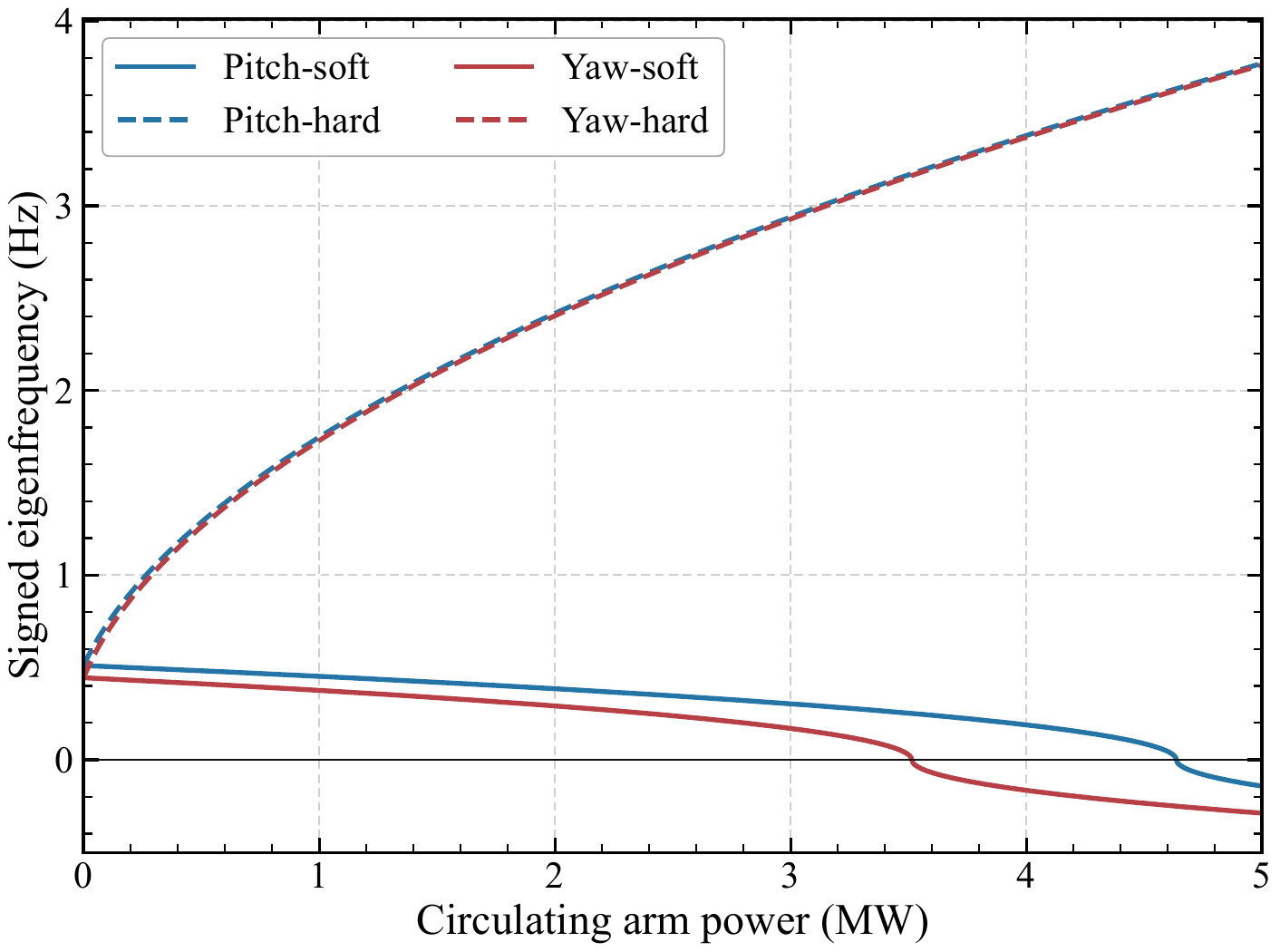}
    \caption{Frequency of the hard and soft modes for pitch and yaw in the arm cavities, as a function of the intra-cavity power.}
    \label{fig:AngularFreq}
\end{figure}

At megawatt circulating power, angular instabilities become an important limitation of the interferometer. In particular, radiation pressure couples the angular motion of the test masses to the optical mode structure, giving rise to the well-known Sidles--Sigg angular instabilities~\cite{Sidles_06}. The radiation-pressure torque can either soften or stiffen the mechanical angular modes, depending on the optical configuration, and can overcome the mechanical restoring torque at sufficiently high power.

Angular stability is determined by the balance between the mechanical restoring torque and the radiation-pressure torque. Importantly, the mechanical restoring torque scales as $\sim M_{\rm tm}^{5/3}$, while the radiation-pressure torque scales as $\sim P_{\rm arm}L$. Compared with the LIGO and Virgo detectors, the mechanical restoring torque increases by a factor of $\approx 54$, assuming the same angular-mode eigenfrequencies for the test masses, whereas the radiation-pressure torque increases by a factor of $\approx 14$. Therefore, optical rigidity is expected to be less severe in the proposed UI configuration than in existing GW detectors.

To maintain a stable cavity at \SI{3}{\mega\watt} of circulating power, we exploit both the large inertia of the heavy test masses and an ``ear'' geometry that increases the separation between the suspension fibres. The increased fibre separation provides a longer lever arm and therefore enhances the mechanical restoring stiffness in both pitch and yaw. This design also results in higher angular eigenfrequencies of the suspension and increased coupling of ground vibrations compared with a softer suspension. However, the angular noise level is suppressed by the active platform, as discussed in Sec.~\ref{sec:AngularControl}. The complete test-mass geometry is shown in Fig.~\ref{fig:TestMass}.

The resulting suspension restoring torque coefficient, $k_{\mathrm{res}}$, together with the radiation-pressure torque coefficients for the soft and hard angular modes, $k_S$ and $k_H$, are given in Table~\ref{tab:restoring_rp_torque}. The resonant frequencies of the soft and hard modes in pitch and yaw as a function of arm-cavity power are shown in Fig.~\ref{fig:AngularFreq}. The cavity remains stable up to \SI{3.5}{\mega\watt}.

\begin{table}[htbp]
\centering
\caption{Suspension restoring torque and radiation-pressure torque at a circulating power of 3\,MW.}
\label{tab:restoring_rp_torque}
\begin{tabular}{lrrrr}
\toprule
DOF & $k_{\rm res}$ & $I$ & $\kappa_{\rm soft}$ & $\kappa_{\rm hard}$\\
& (\si{\newton\metre\per\radian})
& (\si{\kilogram\metre\squared})
& (\si{\newton\metre\per\radian})
& (\si{\newton\metre\per\radian})\\
\midrule
Pitch & 236.3 & 23.0 & $-153.1$ & 7605.6\\
Yaw   & 178.9 & 23.0 & $-153.1$ & 7605.6\\
\bottomrule
\end{tabular}
\end{table}



\subsection{Angular controls}\label{sec:AngularControl}

In addition to the longitudinal degrees of freedom, seismic-induced angular motion must also be controlled. 
Previous studies have identified angular sensing and control as a potential limitation to ET sensitivity below 20\,Hz \cite{MaggioreAngularControlsNoise} and proposed inertial sensors to tackle the problem of inertial isolation~\cite{Ding2022_InterferometricInertialSensor, Collette2015_InterferometricAbsoluteMotion, Amorosi2022_CompactVertical, Viets2018WattLinkage, vanHeijningen2022Cryogenic, Sider2023ETEST}.

To mitigate the angular controls problem, we introduce active platform stabilization as a pre-isolation stage upstream of the 6-stage suspension chains. The Compact-6D seismometer \cite{6Dproposal,6Dproto,6Dsilica} is well suited to this application, having demonstrated six-degree-of-freedom platform stabilization\,\cite{6Dplatform} and longitudinal isolation from 10\,mHz\,\cite{6DPRL}. 

\begin{figure}[t]
    \centering
    \includegraphics[width=\columnwidth]{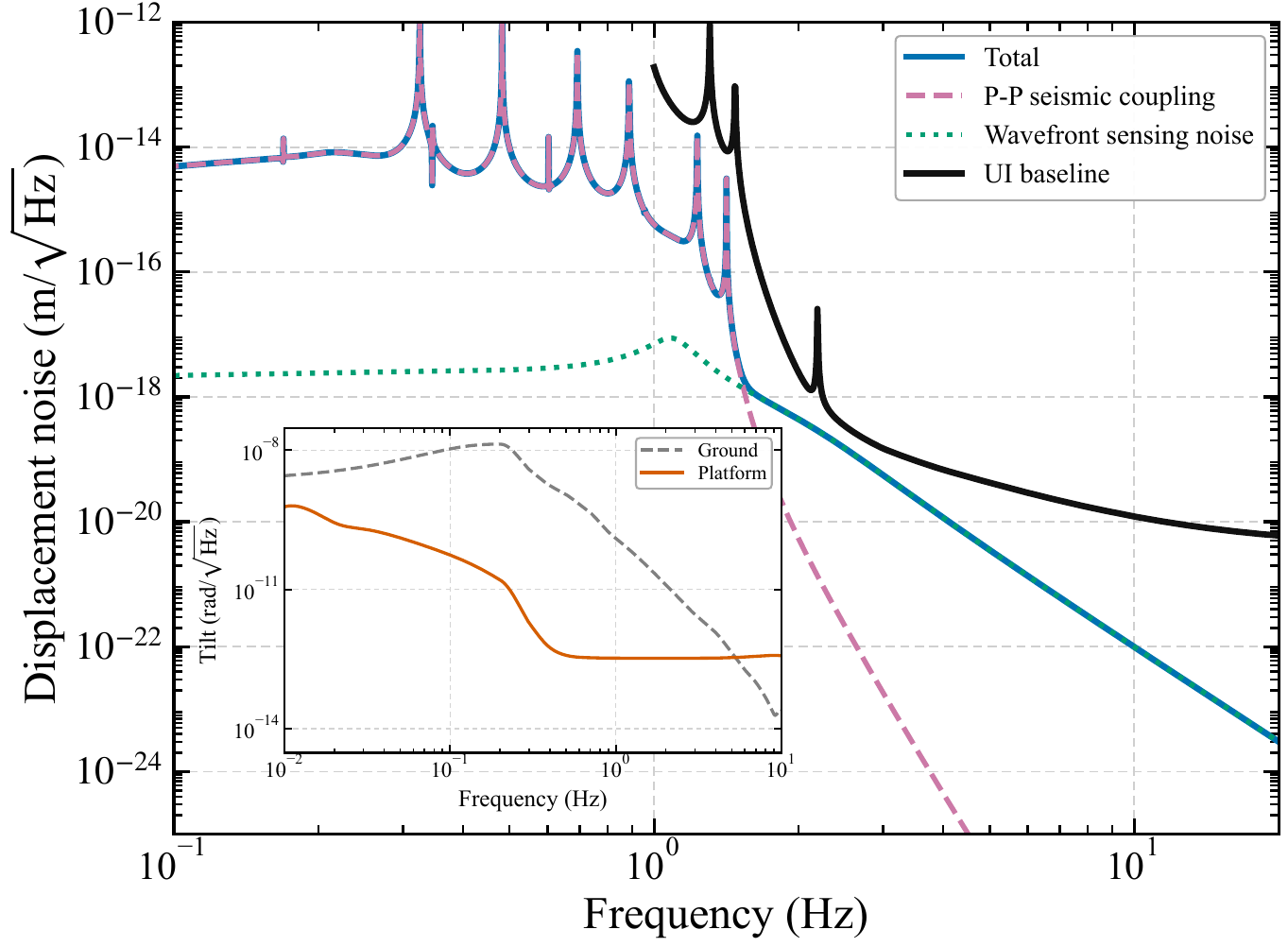}
    \caption{Seismic and sensor noise coupling of the angular motion to the proposed detector. A control bandwidth of 0.7\,Hz keeps the wavefront-sensing noise below the target sensitivity in the detection band.}
    \label{fig:ETangular}
\end{figure}

The resulting low-frequency isolation provides at least two orders of magnitude of suppression of the tilt motion in the critical \SIrange{0.1}{1}{\hertz} band as shown in the inset of Fig.~\ref{fig:ETangular}. This reduction in support motion and the use of wavefront-sensing and control
mitigates seismic-induced angular motion below the sensitivity requirement of UI. In addition, the angular control can further improve alignment stability and enhance detector robustness\,\cite{VirgoAlignment1,VirgoAlignment2}. We implement a 0.7\,Hz angular control bandwidth, sufficiently low to prevent wavefront-sensing noise from contaminating the detection band, where the wavefront-sensing noise is assumed at the level of $10^{-15}\,\mathrm{rad/\sqrt{Hz}}$. 


\subsection{Optical clipping}
\label{sec:test-mass}

Having targeted the circulating arm power at \SI{3}{\mega\watt} and the
beam radius at \SI{13.5}{\centi\meter}, we now consider the corresponding
test-mass geometry. The mirror aperture must be sufficiently large to
keep the effective loss of the fundamental arm-cavity mode negligible.
For a Gaussian beam incident on a circular mirror, the fraction of power
that lies outside the mirror radius is
\begin{equation}
    l_{\mathrm{c}}
    =
    \exp\left(-\frac{2r^{2}}{w^{2}}\right),
\end{equation}
where $r$ is the mirror radius and $w$ is the beam radius.

Considering the diffraction effect, the amplitude coefficient
obtained by projecting the remaining field back onto the original
fundamental mode is
\begin{equation}
    c_{00}
    =
    1-l_{\mathrm{c}}.
\end{equation}
The fraction of power retained in the resonant fundamental mode is therefore $\lvert c_{00}\rvert^{2}$, and the corresponding effective single-mirror loss is
\begin{equation}
    l_{\mathrm{eff}}
    =
    1-\lvert c_{00}\rvert^{2}
    =
    1-\left(1-l_{\mathrm{c}}\right)^{2}
    \simeq 2l_{\mathrm{c}}.
\end{equation}
This effective loss contains both the power directly removed by the finite aperture and the power scattered from the fundamental mode into higher-order spatial modes (HOMs). The corresponding round-trip effective loss becomes
approximately $2l_{\mathrm{eff}}\simeq4l_{\mathrm{c}}$.
We select the test mass radius of \SI{37}{\centi\meter} and thickness of \SI{46.5}{\centi\meter}. With the adopted beam radius of \SI{13.5}{\centi\meter}, the direct clipping loss is approximately $0.3$\,ppm per mirror, while the corresponding
effective loss of the resonant fundamental mode in the cavity is approximately $1.2$\,ppm. The selected mirror therefore provides sufficient aperture for the fundamental arm-cavity mode while retaining margin for alignment
offsets.

For amorphous coating systems, large-area ion-beam-sputtering (IBS) coating is already an established technology within the gravitational-wave community. The LMA coating facility, developed for the Virgo and LIGO programmes, has demonstrated low-loss coatings on large interferometer optics~\cite{Cimma_06}, while the LMA Grand Coater was designed to accommodate substrates up to \SI{800}{\milli\metre} in diameter~\cite{VirgoFutureVision_2016}. Earlier development of the Virgo coating facility also demonstrated the feasibility of IBS coating systems for optics up to \SI{700}{\milli\metre} in diameter~\cite{Vinet_04}. The existing coating system at the University of Strathclyde is designed to coat optics up to \SI{620}{\milli\metre} in diameter and \SI{200}{\kilogram} in mass, with the capability to coat two optics simultaneously~\cite{Strathclyde_IBD_2020}. More recently, larger-area IBS coating capability has been developed at the Laser Zentrum Hannover (LZH), including a system designed for optics up to \SI{2}{\meter} in diameter~\cite{Kellermann_25,Badorreck_25}. 
The challenge is also not unique to the proposed ET design: Cosmic Explorer plans to employ test masses of comparable mass and size~\cite{Evans_22}, making the development of suitable large-scale coating infrastructure a common requirement for next-generation gravitational-wave observatories.
This shared requirement provides an opportunity for coordinated development of coating technologies and facilities across future observatories.

\subsection{Parametric instabilities}

The heavy test mass is beneficial at low frequencies; however, it also brings a challenge associated
with its internal acoustic modes. At high circulating power, an acoustic mode can scatter carrier light from the fundamental optical mode into an HOM. If the acoustic-mode frequency is sufficiently close to the corresponding optical-mode separation, the resulting radiation-pressure beat pattern can drive the same mechanical mode and produce parametric instability\,\cite{Braginsky2001PI,PhysRevLett.114.161102}. If the mechanical mode frequency is low, it can also contaminate the sensitivity appearing as a narrow thermal-noise
feature in the kilohertz band\,\cite{PhysRevD.52.577}. Preliminary finite-element calculations for the selected test-mass
geometry give the lowest internal acoustic mode at approximately
\SI{3.36}{\kilo\hertz}.  

To suppress the parametric instabilities and acoustic-mode thermal noise, we could actively damp the selected modes using a resonant transducer attached to the test mass. The concept is adapted from the mechanical displacement transformers used in resonant-mass gravitational-wave detectors. In these systems, a lightweight secondary resonator, often implemented with a mushroom-like geometry, is tuned to
a mode of the primary mass\,\cite{PhysRevD.54.1264,10.1063/1.1544077,
Gusev_2008}. Near the coupled resonance, the motion of the secondary resonator provides an enhanced local signal that can be sensed with an interferometric readout. For the proposed detector, a mushroom-shaped transducer could be tuned to the approximately \SI{3.36}{\kilo\hertz} lowest internal acoustic
mode. A compact local interferometer~\cite{6DPRL, Cooper2018HoQI, Isleif2019CompactMultifringe} would measure the transducer motion and provide an error signal for feedback actuation, thereby reducing the effective mechanical quality factor of the unstable mode. Active damping
of a parametrically unstable test-mass mode has already been demonstrated in Advanced LIGO using electrostatic actuation\,\cite{PhysRevLett.118.151102,pan2026demonstrationopticalfeedbackcontrol}. The resonant-transducer approach considered here combines this active-damping principle with mechanical displacement
enhancement, potentially reducing the required sensing and actuation strength while avoiding the broadband mechanical dissipation associated with a conventional passive damper.  Optical feedback provides a complementary approach and has been demonstrated first in an \SI{80}{\metre} cavity and, more recently, in Advanced LIGO, suppressing the instability through destructive interference with the responsible higher-order optical mode\,\cite{PhysRevD.109.102006,pan2026demonstrationopticalfeedbackcontrol}.

\section{Sensitivity analysis}

In this section, we discuss the detector parameters optimised for broadband, low-frequency and kHz operation, with the goal of achieving optimal sensitivity to the inspiral phase of compact-binary mergers and exploring the physics of neutron stars.
The achievable sensitivity is ultimately constrained not only by the interferometer itself but also by the observatory infrastructure~\cite{DiGiovanni2020SosEnattos, Koley2022Terziet}.

Apart from the gravity-gradient noise considered in Sec.~\ref{sec:sus}, residual gas noise is another facility-level contribution to the overall sensitivity.
Residual gas adds noise through both gas damping of the test masses at low frequencies and broadband phase noise caused by molecules traversing the laser beam~\cite{CAVALLERI20103365}. We include these effects using the residual-gas model adopted for the ET design, with the vacuum pressure and gas composition chosen to represent the expected operating conditions of the proposed facility. The assumed partial pressures of residual H$_2$, N$_2$, H$_2$O, and O$_2$ in the beam tubes are \SI{1e-8}{\pascal}, \SI{2.5e-10}{\pascal}, \SI{5e-9}{\pascal}, and \SI{2.5e-10}{\pascal}, respectively. The corresponding partial pressures in the vacuum chambers are \SI{2e-8}{\pascal}, \SI{5e-10}{\pascal}, \SI{1e-8}{\pascal}, and \SI{5e-10}{\pascal}.

\subsection{Quantum noise}
\label{sec:arm_power}

\begin{figure}[t]
    \centering
\includegraphics[width=\columnwidth]{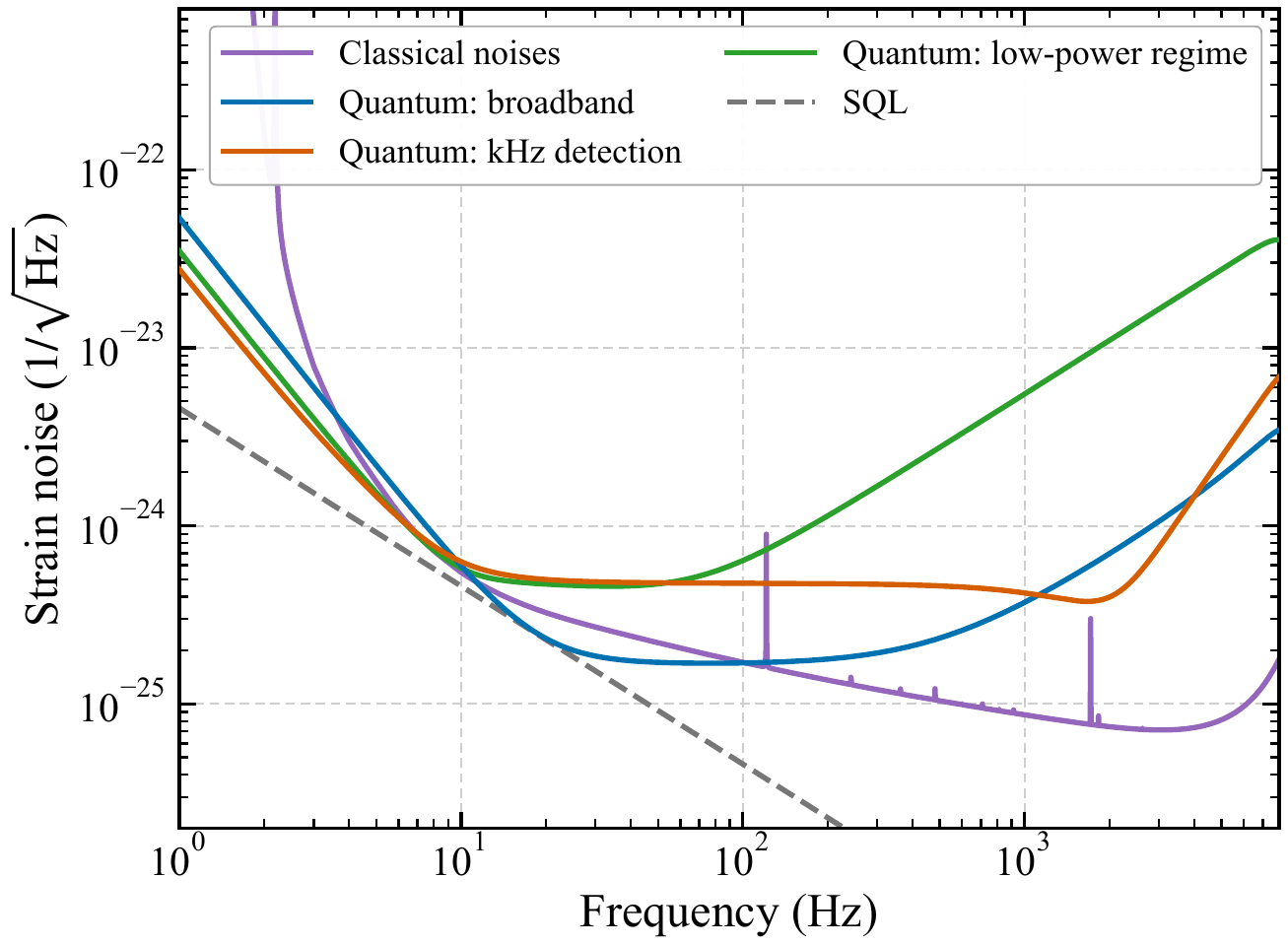}
    \caption{
    Quantum noise compared with the total classical noise and the squeezed standard quantum limit (SQL) for arm powers and GW signal gains optimised for broadband sensitivity, kilohertz detection, and low-power operation of UI.
    }
    \label{fig:ETUILP}
\end{figure}

The quantum-noise power spectral density (PSD) of a tuned interferometer without injected squeezed light can be expressed as~\cite{PhysRevD.67.062002}
\begin{equation}
S=\frac{h_{\rm SQL}^2(\omega)}{2}
\left[\frac{1}{\kappa(\omega)}+\kappa(\omega)\right],
\end{equation}
where
\[
h_{\rm SQL}^2(\omega)=\frac{8\hbar}{M_{\rm tm}\omega^2L^2}
\]
is the squared strain standard quantum limit~\cite{BraginskyKhalili1992}. The first and second terms in brackets represent shot noise and quantum radiation-pressure noise, respectively. The optomechanical coupling factor, \(\kappa(\omega)\), is given by
\begin{equation}
    \kappa(\omega)
    =
    \frac{2\Theta\omega_c}
    {\omega^2\left(\omega_c^2+\omega^2\right)},
    \hspace{8mm}
    \Theta=\frac{8\omega_0 P_{\rm arm}}{M_{\rm tm}cL},
    \label{eq:kappa}
\end{equation}
where \(\omega_c\) is the interferometer bandwidth, \(P_{\rm arm}\) is the circulating arm power, and \(\omega_0\) is the laser angular frequency. The total quantum noise reaches the SQL when \(\kappa(\omega)=1\).

The interferometer response is characterised by two poles: the optomechanical pole, \(\omega_q\), and the coupled-cavity optical pole, \(\omega_c\). When these poles are not well separated in frequency, two filter cavities are required to achieve the frequency-dependent rotation of the squeezing angle. In this work, we restrict the implementation of frequency-dependent squeezing to a single filter cavity to simplify the optical configuration and reduce the degradation of squeezing due to optical losses in a second filter cavity. This approach is valid when
\begin{equation}
    \omega_q
    \approx
    \sqrt{\frac{2\Theta}{\omega_c}}
    \ll \omega_c.
    \label{eq:kappa_app}
\end{equation}

In this section, we analyse three interferometer configurations: broadband, high-bandwidth, and low-bandwidth. These are optimised for mid-frequency sensitivity, kilohertz detection, and low-frequency sensitivity at low optical power, respectively, as shown in Fig.~\ref{fig:ETUILP}. For each configuration, we set $\omega_q \sim 2\pi \times 10$\,Hz and ensure that the condition in Eq.~\eqref{eq:kappa_app} is satisfied.

\subsection{Broadband configuration}

The broadband configuration of the detector optimises its sensitivity to black hole and neutron star mergers and chooses $\omega_c \approx 2\pi \times 540$\,Hz to balance this quantum noise with the sum of classical noises in the observatory's most sensitive frequency band from 20\,Hz up to a few hundred Hz. 
Similar to the LIGO and Virgo detectors, the largest classical noise source that determines the choice of the bandwidth, $\omega_c$, is the coating Brownian noise~\cite{PhysRevD.87.082001}. The scaling of the coating thermal noise for a 15-km baseline detector relative to the LIGO-estimated value~\cite{Granata_2020, Martynov2017, Capote2025AdvancedLIGO, Gras2017} is given by
\begin{equation}
\begin{split}
\sqrt{S_{\rm CTN}(100\,{\rm Hz})} &\approx 1.2 \times 10^{-20} \frac{\rm m}{\sqrt{\rm Hz}} \\ &\times \frac{6.2\,{\rm cm}}{w} \sqrt{\frac{\phi_H}{3 \times 10^{-4}}},
\end{split}
\end{equation}
where $w = 13.5$\,cm is the beam size in the UI and $\phi_H = 5\times 10^{-5}$ is the required loss angle of the high-index-of-refraction coating material.
The last two factors in the above equation suppress the coating noise by a factor of 5 compared to the Advanced LIGO level.

In the UI noise budget, we adopt target mechanical loss angles of
$1.3\times10^{-5}$ and $5\times10^{-5}$ for the low- and high-index coating materials, respectively. The latter represents an approximately factor-of-two reduction in the high-index loss angle relative to the
current A\# baseline value~\cite{PhysRevD.105.102008, Davenport:24} but is more conservative than the A\# optimistic coating noise~\cite{PhysRevLett.127.071101, sun2026ligoasharpdetectordesign, collaborationreport}.
For the adopted parameters, coating thermo-optic and substrate thermal noises remain below coating Brownian noise as shown in Fig.~\ref{fig:rt-budget}.

Several candidate materials and coating architectures may offer routes toward this goal, albeit introducing additional technical challenges relating to material, growth method, and scalability. As an example, improvements of \(\mathrm{TiO_2}\):\(\mathrm{GeO_2}\) will likely involve more extensive heat treatment campaigns, which can exacerbate substrate surface figure changes and defect formation in the coating, leading to higher optical loss and scattering. The mixture \(\mathrm{TiO_2}\)--\(\mathrm{SiO_2}\) is another promising high-index material for which post-deposition annealing is a critical step, with blistering and cracking having been observed after heat treatment~\cite{PhysRevLett.131.171401}. Further pathways are amorphous semiconductor and multi-material coatings ~\cite{PhysRevD.91.042002,PhysRevD.91.042001,PhysRevLett.125.011102,8mhb-mmtz,PhysRevD.111.042003}, whose thermal noise performance must be carefully balanced against their optical absorption. Crystalline coatings made from GaAs/AlGaAs can go lower than the requirement of UI\,\cite{cole2013tenfold, cole2023substrate}, although substantial technological scale-up would be required for the large optics considered here.  More generally, coating loss may depend not only on material composition but also on structure across multiple length scales. Structural characterization paired with atomistic modelling has identified specific nm-sized motifs that contribute to the mechanical dissipation, providing key insights to design future coatings\,\cite{PhysRevLett.123.045501}, while recent measurements of ion-beam-sputtered silica films suggest that elastic and structural anisotropy may represent another relevant factor in the mechanical dissipation of amorphous coatings\,\cite{fmrf-dys9}.

\subsection{kHz detection}
\label{sec:HF}

In this section, we consider the detector configuration optimised for high-frequency operation, with particular emphasis on the sensitivity required to resolve the post-merger oscillations of neutron stars. These signals, which lie predominantly in the kilohertz frequency range, carry direct information about the dynamics and thermodynamics of nuclear matter and therefore provide a unique probe of the neutron-star equation of state. Achieving high sensitivity at these frequencies is consequently an important scientific goal for the next generation GW detectors.

We exploit the coupled-cavity resonance formed by the arm and signal-extraction cavities (SEC) to enhance sensitivity in the kHz range~\cite{PhysRevD.99.102004}. The mode-splitting frequency and the bandwidth of each mode can be approximately written as
\begin{equation}
\omega_{c}=\frac{c\sqrt{T_{\rm itm}}}{2\sqrt{L_{\rm arm}L_{\rm sec}}}, \qquad
\gamma_c=\frac{cT_{\rm sem}}{4L_{\rm sec}}\,.
\end{equation}
A resolvable coupled-cavity resonance emerges when $\gamma_c < \omega_c$. The \SI{8}{\percent} signal-extraction-mirror transmissivity defines the broadband operating point. Reducing the transmissivity to \SI{1}{\percent} narrows the coupled-cavity response and enhances its peak sensitivity in the kilohertz range, providing improved sensitivity to neutron-star post-merger signals as shown in Fig.~\ref{fig:AR}.

We find that transitioning between the broadband and kilohertz-optimised configurations requires only the change in the signal-extraction-mirror transmissivity and the filter-cavity input-mirror transmissivity for in-cooperating frequency dependent squeezing. Note that the high-frequency sensitivity is unaffected by changing the filter-cavity input-mirror transmissivity, which influences only the low-frequency squeezing. This opens the possibility of making the interferometer dynamically tunable~\cite{DeVine2002, DeVine2002CQG} between different observing modes rather than requiring a dedicated optical configuration for each science target. One possible implementation is a signal-extraction-mirror with coatings on both sides, designed such that its effective transmissivity can be varied through controlled changes in the mirror-substrate temperature. Similar to the design of the input test masses in Virgo~\cite{Hild_09, Khalili_05}, thermal tuning would provide an actuator for adjusting the mirror transmissivity without replacing the optic, allowing the coupled-cavity response to be moved between a broadband configuration and a narrower, higher-sensitivity response in the kilohertz range.


An important practical limitation on high-frequency sensitivity is optical loss in the signal-extraction cavity. Scattering, mode mismatch, thermally induced wavefront distortions, and higher-order-mode generation all contribute to the effective loss, reducing the signal response and degrading the benefits of squeezed light. In the sensitivity calculation, we assume a total round-trip signal-extraction-cavity loss of $500\,\mathrm{ppm}$.

\subsection{Low-power regime}

In this section, we show that UI can maintain its low-frequency sensitivity in a low-power operating mode. While reducing the circulating arm power, the GW signal gain can be proportionally increased by adjusting the finesse of the signal-recycling cavity to preserve the quantum-noise level of the broadband configuration below the coupled-cavity pole frequency, \(\omega_c\). However, reducing the arm power also lowers \(\omega_c\), potentially bringing it closer to the optomechanical pole frequency, \(\omega_q\). Therefore, Eq.~\ref{eq:kappa_app} imposes a lower limit on the circulating arm power required to maintain UI sensitivity in the \(3\text{--}30\,\mathrm{Hz}\) band.

In this configuration, we choose \(\omega_c \approx 2\pi\times 50\,\mathrm{Hz}\), \(\omega_q \approx 2\pi\times 11\,\mathrm{Hz}\) to satisfy the single-filter-cavity condition. The corresponding circulating arm power is
\begin{equation}
P_{\rm arm} \approx \frac{\omega_c}{\omega_q} \times \frac{M_{\rm tm}cL}{16\omega_0}\omega_q^3
\approx 105\,\mathrm{kW},
\end{equation}
with a required input power of \(17.4\,\mathrm{W}\) and a signal-recycling mirror transmissivity of 60\,\%.

This low-power configuration allows UI to approach the low-frequency infrastructure noise limit for a representative ET facility at Sos Enattos, while addressing quantum radiation-pressure noise using \(\SI{440}{\kilogram}\) test masses without detuning the signal-recycling cavity. It also mitigates the risk of technical low-frequency noise that may arise during high-power operation in the broadband configuration.

\subsection{Sensitivity scaling with the arm length}

Arm length is a key parameter in the system-level design of a gravitational-wave detector. For displacement noise with a physical amplitude spectral density that is independent of the arm length \(L\), the corresponding strain noise scales as \(L^{-1}\). For a fixed interferometer bandwidth, with the circulating arm power and test-mass mass held constant, the strain-referred quantum shot noise scales approximately as \(L^{-1/2}\), whereas quantum radiation-pressure noise scales as \(L^{-3/2}\). Coating thermal noise scales as \(L^{-1}\) for a fixed beam size, or approximately as \(L^{-3/2}\) if the arm-cavity \(g\) factor is held constant. 

\begin{figure}[t]
    \centering
    \includegraphics[width=1\columnwidth]{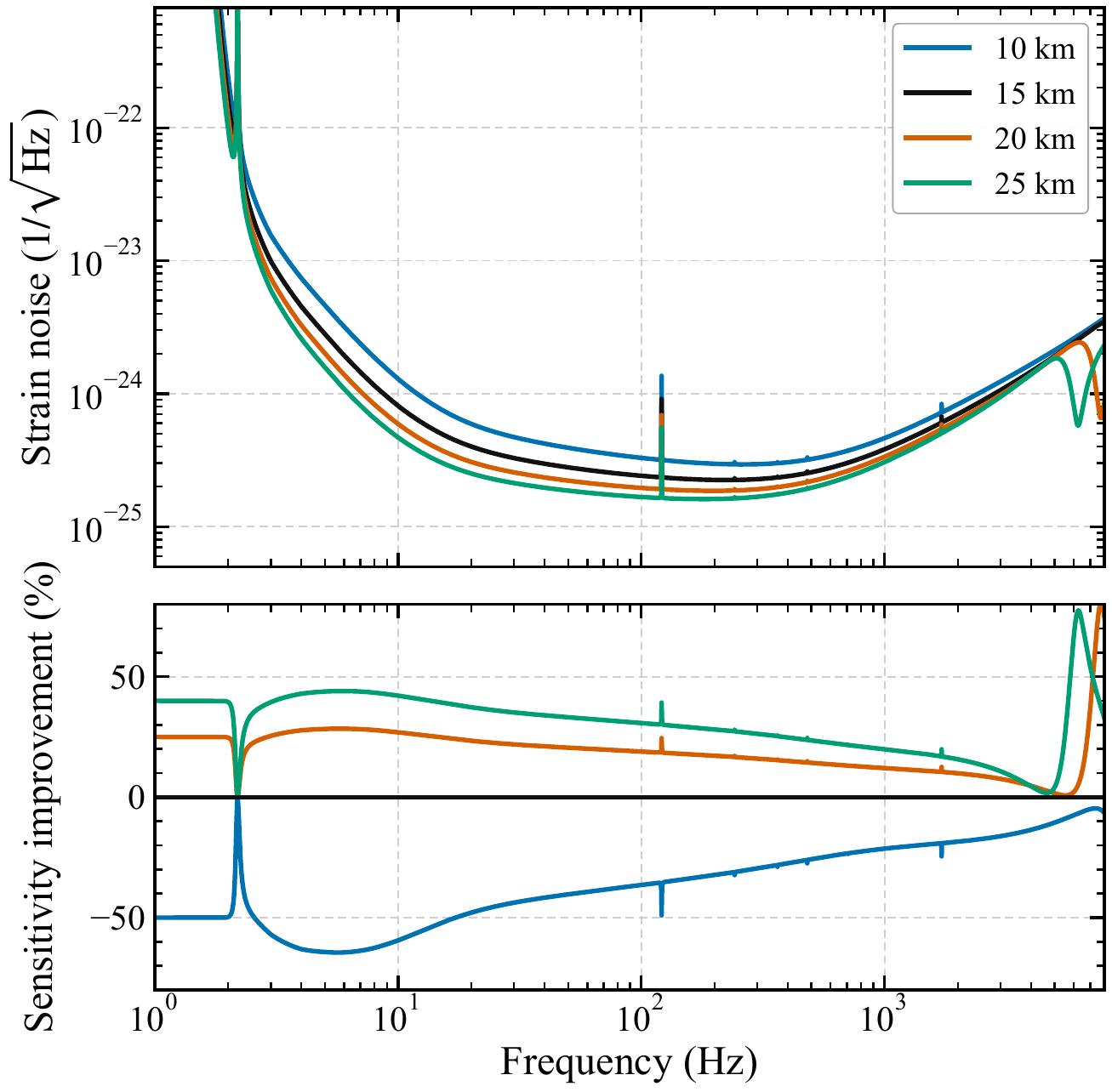}
    \caption{Comparison of strain sensitivities for different interferometer arm lengths. The upper panel shows the sensitivity for arm lengths of \SI{10}{\kilo\meter}, \SI{15}{\kilo\meter}, \SI{20}{\kilo\meter}, and \SI{25}{\kilo\meter}. The lower panel shows the sensitivity improvement relative to the \SI{15}{\kilo\meter} reference design, with positive values indicating lower strain noise.}
    \label{fig:ArmLength}
\end{figure}

In this section, we compare designs with arm lengths of \SI{10}{\kilo\meter}, \SI{15}{\kilo\meter}, \SI{20}{\kilo\meter}, and \SI{25}{\kilo\meter}. The circulating arm power and input test mass transmissivity are held fixed, while the signal extraction mirror transmissivity is adjusted for each arm length to maintain the same effective interferometer bandwidth as the \SI{15}{\kilo\meter} reference design. We keep the beam radius, rather than the arm-cavity \(g\) factor, fixed because further increasing the beam size is constrained by the finite test-mass aperture. The resulting sensitivities for the broadband configuration are shown in Fig.~\ref{fig:ArmLength}. 

We find that increasing the arm length produces the most pronounced improvement at low frequencies, where quantum radiation-pressure noise is a dominant contribution in the few-hertz band. At higher frequencies, the scaling depends on the dominant noise source. For shorter arm lengths, where coating thermal noise limits the sensitivity, the strain noise improves approximately as $L^{-1}$ from 15\,km up to 20\,km. As the arm length increases and quantum shot noise becomes dominant, the improvement becomes weaker, scaling approximately as $L^{-1/2}$ for arm lengths above $\approx 20$\,km. The precise transition between these two scaling regimes depends on the coating loss angle and circulating arm power.


\section{Comparison with the Cosmic Explorer design}

\begin{figure}[t]
    \centering
    \includegraphics[width=1\columnwidth]{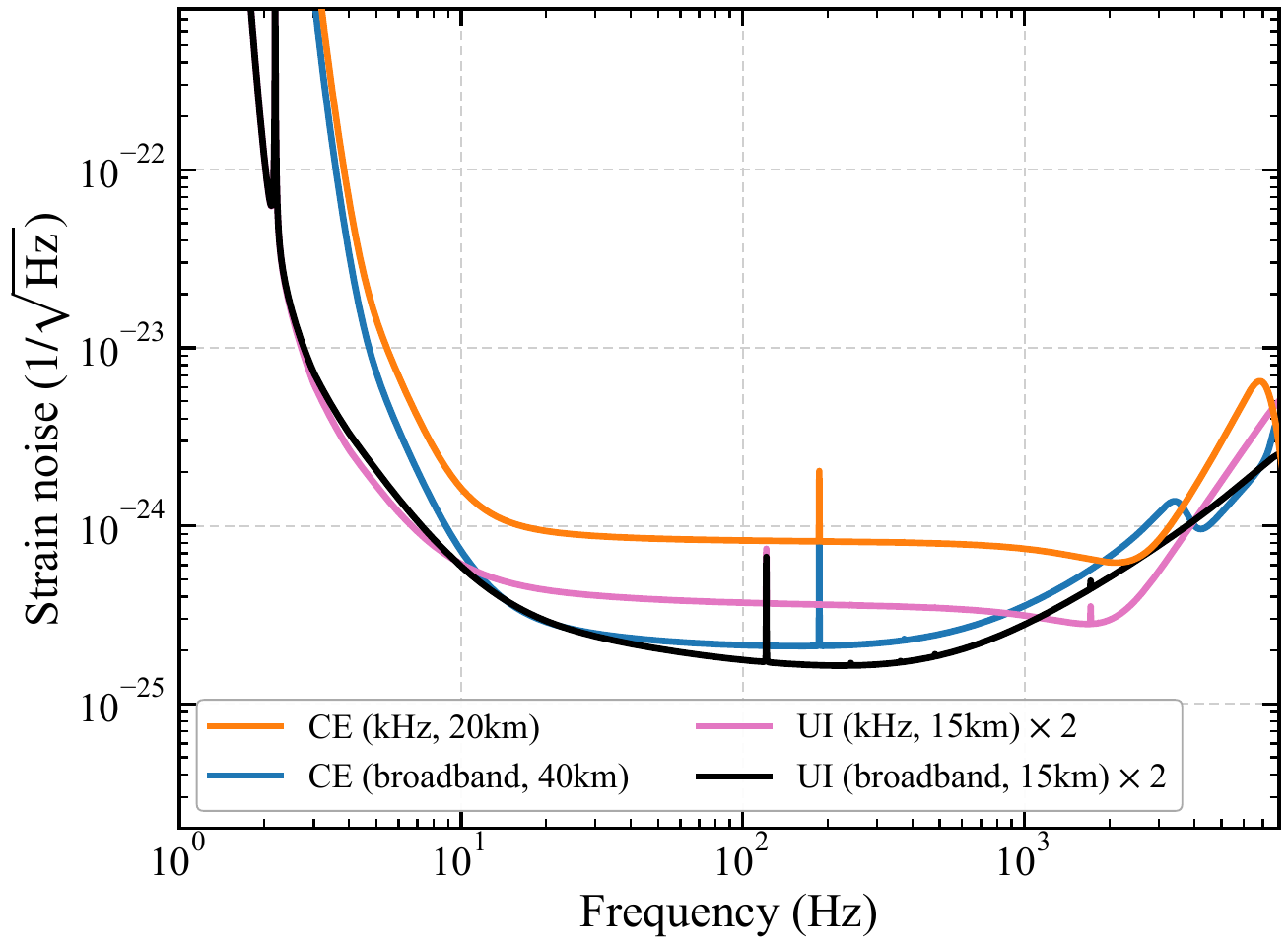}
    \caption{Comparison of the sensitivities of the broadband and kilohertz-targeted CE and ET configurations. The ET curves represent the network sensitivities of two \SI{15}{\kilo\meter} L-shaped interferometers rotated by $45^{\circ}$ with respect to each other. All sensitivities correspond to the antenna response  for a source at a zenith angle of
$15^\circ$.}
    \label{fig:AR}
\end{figure}

\begin{figure}[tbp]
  \centering
  \includegraphics[width=\columnwidth]{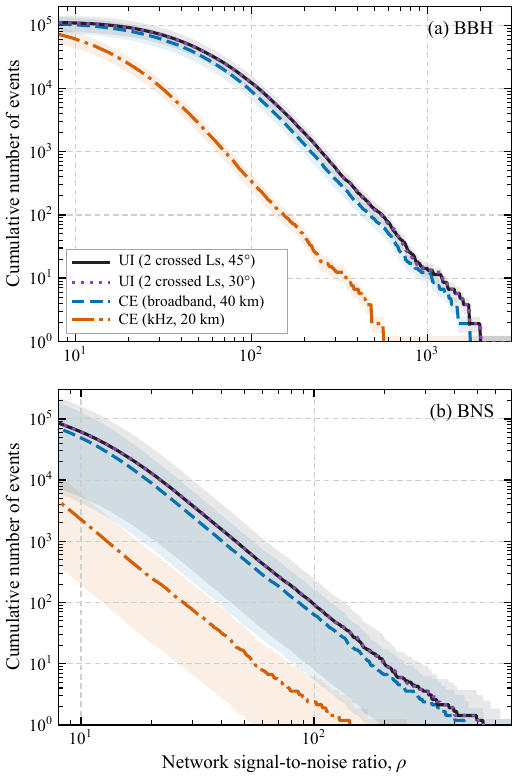}
  \caption{Cumulative number of compact-binary events in one year with network
  SNR greater than $\rho$ for (a) BBHs and (b) BNSs.  The black solid curve and purple dotted line are
  the proposed ET 15\,km broadband configuration, consisting of two L-shaped interferometers rotated by $45^\circ$ and $30^\circ$ relative to one
  another respectively; the blue dashed and orange dash-dotted curves are, respectively, a
  single CE 40~km broadband interferometer and a single CE 20~km
  high-frequency interferometer.  SNRs are calculated from the public one-year
  catalogs of Ref.~\cite{Branchesi2023ETDesigns}, using
  \textsc{IMRPhenomXHM} for BBHs\,\cite{Garcia-Quiros:2020qpx} and
  \textsc{IMRPhenomXAS\_NRTidalv3} for BNSs\,\cite{Pratten:2020fqn,Abac:2023ujg}, with a 2~Hz low-frequency
  cutoff, the full antenna response, the rotation of the Earth, and continuous
  operation.  The BBH central curves use
  $R_{\rm BBH}=38.45\,\mathrm{Gpc}^{-3}\,\mathrm{yr}^{-1}$, the arithmetic
  midpoint of the GWTC-5.0 joint 90\% interval $27.5$--$49.4$ at $z=0.2$; the
  BNS central curves use the GWTC-5.0 FullPop median
  $R_{\rm BNS}=59.3\,\mathrm{Gpc}^{-3}\,\mathrm{yr}^{-1}$ at $z=0$, and the
  BNS bands span the joint 90\% interval $5.1$--$154.7$ (all rates in
  $\mathrm{Gpc}^{-3}\,\mathrm{yr}^{-1}$)\,\cite{LVK2026GWTC5Population}.
  Shaded regions show merger-rate uncertainty only.}
  \label{fig:compact-binary-cumulative-snr}
\end{figure}


The proposed detector shares several key design features with the Cosmic Explorer \SI{40}{\kilo\meter} and \SI{20}{\kilo\meter} concept~\cite{Evans2021CEHorizon}, including room-temperature operation based on fused-silica technology and an optical configuration that retains strong sensitivity to neutron-star merger
signals in the kilohertz band \cite{PhysRevD.99.102004}. The underground environment of ET, however, provides a substantially lower environmental noise floor. In particular, reduced ground motion and Newtonian noise make it possible to extend the detector sensitivity to
lower frequencies.

The enhanced sensitivity of the two crossed 15\,km UI detectors for ET compared with a single 40\,km Cosmic Explorer detector, as shown in Fig.~\ref{fig:AR}, can be understood, to first order, by considering the arm lengths and circulating optical powers. Similar to ET-HF, the proposed UI design assumes 3\,MW of circulating power in each arm, compared with a more conservative baseline of 1.5\,MW adopted by Cosmic Explorer. The ultimate circulating optical power is expected to be limited by optical absorption in the mirror coatings, including spatially non-uniform absorption that can lead to thermal lensing and wavefront distortions \cite{Brooks2006, LIGO2021PointAbsorbers}.
We note, however, that achieving higher circulating powers in longer interferometers may present additional technical challenges, for example due to the negative optical rigidity discussed in Sec.~\ref{sec:OpticalRigidity}. The factor of 2.67 in arm length between a 40-km Cosmic Explorer and UI for ET does not lead to a linear scaling in sensitivity, but rather an improvement approximately proportional to $\sqrt{L}$, as discussed in the following section. This factor is also partially compensated, by a factor of approximately 1.3, by the use of two crossed UI detectors for ET. Increasing the arm length of UI for ET from the baseline 15\,km to 20\,km would further reduce the risk associated with the assumptions made regarding the arm power and coating thermal noise.

To quantify their performance in astrophysics, we compute
the cumulative annual yield, $N(>\rho)$, of compact-binary signals with network
signal-to-noise ratio (SNR) larger than $\rho$.  We retain the population shapes
and one-year realizations of the public BBH and BNS catalogs constructed for the
ET design-comparison study\,\cite{Branchesi2023ETDesigns}, and calculate the SNRs
with \textsc{GWFAST}\,\cite{Iacovelli2022GWFAST}, including the source-dependent
antenna response and the rotation of the Earth.  We adopt a low-frequency cutoff
of 2~Hz and the catalog luminosity distances based on\,\cite{Planck2020Cosmology}.  The ET curve represents the quadrature
network SNR of two L-shaped interferometers with 15~km arms
and a relative rotation of $45^\circ$; the CE 40~km broadband and CE 20~km
high-frequency curves each represent a single L-shaped interferometer, using the
corresponding CE design sensitivities\,\cite{Evans2021CEHorizon}.  The comparison
is therefore between the specified observatory configurations, rather than
between equal numbers of interferometers.  We
update the overall population normalizations to the GWTC-5.0 merger-rate
constraints\,\cite{LVK2026GWTC5Population}; the shaded intervals in
Fig.~\ref{fig:compact-binary-cumulative-snr} propagate only
this rate uncertainty, while leaving the catalog mass, spin, and redshift
distributions fixed.

For BBHs, the upper panel of Fig.~\ref{fig:compact-binary-cumulative-snr} shows that the crossed broadband
configuration of UI for ET produces the largest cumulative yield throughout the plotted SNR
range.  At the reference cut $\rho\geq 12$, used in the ET design-comparison
study as the entry criterion for parameter-estimation forecasts, the central
expectation is $1.06\times10^{5}$ events per year, compared with
$9.8\times10^{4}$ for CE 40~km broadband and $5.1\times10^{4}$ for CE 20~km
high-frequency.  The distinction is stronger in the scientifically valuable
loud tail: the expected numbers above $\rho=100$ are $1.2\times10^{4}$,
$9.1\times10^{3}$, and $3.4\times10^{2}$, respectively, while above
$\rho=200$ they are $2.0\times10^{3}$, $1.3\times10^{3}$, and approximately
$48$.  The proposed room-temperature configuration therefore lies in the
high-SNR regime identified as central to the ET science case\,\cite{Maggiore2020ScienceCase,Branchesi2023ETDesigns}.  Such a large sample of
loud BBH signals can support precise measurements of masses and spins, the
resolution of subdominant multipoles, studies of formation channels and their
evolution with redshift, and precision tests of the merger and ringdown in the
strong-field regime.  The cumulative distribution is particularly informative
for these applications: once the total number of detections approaches the
underlying annual population, further scientific gain is driven mainly by the
number of events with very large SNR.

The BNS result in the bottom panel of Fig.~\ref{fig:compact-binary-cumulative-snr} gives a complementary test
of both the low- and high-frequency performance.  For the GWTC-5.0 central rate,
the 15~km crossed-L configuration yields $4.4\times10^{4}$ BNS signals per year
above $\rho=12$, compared with $3.4\times10^{4}$ for CE 40~km broadband and
$1.3\times10^{3}$ for CE 20~km high-frequency.  Using $\rho\geq30$, the
``golden-event'' benchmark adopted for BNSs in Ref.~\cite{Branchesi2023ETDesigns},
the corresponding yields are $3.9\times10^{3}$, $2.6\times10^{3}$, and
approximately $76$; above $\rho=100$ they are approximately $93$, $63$, and
$2.4$.  Thousands of loud inspirals would provide a powerful ensemble for
measuring tidal deformabilities and constraining the neutron-star equation of
state, while the best-localized events could be combined with electromagnetic
counterparts or galaxy information for multi-messenger studies and standard
sirens\,\cite{Maggiore2020ScienceCase,Branchesi2023ETDesigns}.  Sensitivity from a
few hertz allows a BNS to be followed for hours, so the time-dependent antenna
response generated by the Earth's rotation can add localization information;
the two crossed L-shaped interferometers additionally provide complementary
polarization response.  At the other end of the band, the proposed detector's
separate kilohertz-tuned operating point targets the tidal and post-merger
signals that carry direct information about hot, dense matter\,\cite{Yang:2017xlf,PhysRevD.99.102004}.  The cumulative-SNR curve shown here uses the
broadband configuration and an inspiral--merger waveform without a post-merger
signal, and therefore does not itself quantify that kilohertz science case.

For comparison, Fig.~\ref{fig:compact-binary-cumulative-snr} also shows the cumulative event counts for a configuration in which the two L-shaped interferometers are rotated by $30^\circ$ relative to one another. The predicted rates differ by less than $1\%$ from those of the $45^\circ$ configuration. Previous studies have shown that the relative orientation of the interferometers can significantly affect their sensitivity to a stochastic gravitational-wave background \cite{Ebersold:2024hgp,Christensen:2026yyl}. The $30^\circ$ configuration therefore represents a promising alternative to enhance stochastic-background science while preserving essentially the same compact-binary detection yield.

\section{Conclusion}
 
We have presented a \SI{15}{\kilo\meter} room-temperature interferometer concept for the Einstein Telescope with the aim of exceeding its broadband sensitivity with lower complexity. The concept exploits the quiet underground environment of ET and proposes heavier test masses, low-loss and strongly dissipation-diluted suspensions, optimised quantum-noise reduction, and enhanced seismic isolation. The proposed configuration employs \SI{440}{\kilogram} fused-silica test masses, \SI{4}{\meter}-long high-stress fused-silica suspension fibres, and a \SI{5}{\kilo\meter} filter cavity to preserve low-loss frequency-dependent squeezing. The intermediate-frequency sensitivity is maintained through a large beam radius and low-loss coating materials, while the high-frequency performance retains the \SI{3}{\mega\watt} circulating power of the ET-HF design.

The concept can achieve the low-frequency baseline sensitivity even with a 100\,kW arm power and can also be optimised for high-frequency GW signals by tuning the transmission of the signal-extraction mirror. The coupled-cavity resonance of the signal-extraction cavity enhances sensitivity in the kilohertz regime and provides a promising route to improved measurements of neutron-star post-merger dynamics and the nuclear equation of state. The signal-extraction mirror can also be made tunable, allowing the detector response to be adjusted between low-frequency, broadband and kHz-optimised operating modes without changing the core interferometer.

We have also investigated the dependence of the achievable sensitivity on arm length. Increasing the arm length provides the largest improvement at low frequencies, where quantum radiation-pressure noise is important, while the gains at higher frequencies depend on the transition between coating thermal noise and quantum shot noise. For the parameters considered here, arm lengths of \SIrange{15}{20}{\kilo\meter} mark an important regime in which further increases in length provide progressively smaller broadband sensitivity improvements.

Our results indicate that the broadband performance targeted by ET does not necessarily require the separation of low- and high-frequency optimization into two interferometers. A single room-temperature interferometer can instead redistribute the noise budget across the frequency band and exploit the advantages of the underground environment, while retaining the mature high-power technologies developed for current-generation detectors.

\section{Acknowledgements}
We acknowledge the members of the Belgian delegation, led by Prof. Giacomo Bruno, who visited the University of Birmingham in June 2026, for the valuable discussions that helped initiate this work. We acknowledge Jan Harms for reviewing this paper. We acknowledge Helios Vocca, Flavio Travasso, Piero Chessa, and Aniello Grado for useful discussions on the suspension thermal noise. We also acknowledge Massimo Granata, Silvia Corezzi, Nelson Christensen and the LIGO Suspension Working Group for useful discussions, and the support of the Institute for Gravitational Wave Astronomy at the University of Birmingham, STFC Consolidated Grant
“Astrophysics at the University of Birmingham” (No.
ST/S000305/1) and UKRI “The next-generation
gravitational-wave observatory network” project (Grant
No. ST/Y00423X/1). H.M. would
like to acknowledge the support from the National Key
R\&D Program of China “Gravitational Wave Detection”
(Grant No.: 2023YFC2205800), and support from Quantum Information Frontier
Science Center. H.Y. is supported by the Natural Science Foundation of China (Grant 12573048).
A.V. acknowledges the support of the Royal Society and Wolfson Foundation.

\bibliography{bibliography}

\end{document}